# The origin of carotenoid triplets in purple photosynthetic bacteria

Juan J. Romero*, Andrew Gall, Viola D'mello, Cristian Ilioaia, Andrew A. Pascal, Bruno Robert*, Manuel J. Llansola-Portoles*

*Université Paris-Saclay, CEA, CNRS, Institute for Integrative Biology of the Cell (I2BC), 91190 Gif-sur-Yvette, France.*

**Abstract**

Photosynthetic antenna proteins harvest light energy while at the same time protecting the organism against photodamage. Carotenoid molecules are essential in the latter process, efficiently quenching unwanted (bacterio)chlorophyll excited states created after photon absorption. (Bacterio)chlorophyll triplets, formed by inter-system crossing, are particularly significant, since in the absence of carotenoid quenching, they sensitise the highly oxidative singlet oxygen. In light-harvesting complex 2 (LH2) from *Rhodoblastus acidophilus*, the pathways that populate carotenoid dark and triplet states remain controversial, involving bacteriochlorophyll-to-carotenoid triplet-triplet transfer and/or generation of triplets by the carotenoid molecules themselves through singlet fission. Transient absorption has been central to understanding photoprotection in these organisms, but spectral congestion limits the separation of the overlapping species needed to discriminate between these pathways. By applying femtosecond stimulated resonance Raman spectroscopy (FSRRS) in different resonance conditions to this protein, in combination with an extension of global analysis to four dimensions (wavenumber, time, intensity and resonance condition), we separate each component of the carotenoid dark-state manifold together with its kinetics. An entangled triplet pair $S^*/^1(TT)$ is observed, which lives about 60 ps, some eight times longer than in solution. However, this stabilisation does not open a pathway to separated triplets or to carotenoid-BChl a heterofission. Triplet–triplet transfer from bacteriochlorophyll *a* to carotenoid is also resolved under BChl a excitation, and fits cleanly as a single 2100 ps component. The carotenoid triplet in LH2 is thus produced by photoprotective triplet-triplet transfer from BChl a, and not by singlet fission.

## Introduction

Photosynthetic antennae capture sunlight and funnel excitation to reaction centres (RCs) with high quantum efficiency, through dense networks of (bacterio)chlorophylls (BChl) and carotenoids precisely positioned within protein scaffolds. This efficiency emerges from a finely-tuned interplay of multiple excited-state pathways, leading the excitation energy to the RCs in tens of picoseconds. At the same time, photosynthetic light-harvesting proteins may be implicated in photoprotective roles, via quenching of unwanted excited states before they can cause damage to the photosynthetic system. The success of photosynthesis relies on this delicate balance between energy transfer and photoprotection. Carotenoid molecules are generally involved in this interplay, as they harvest the energy of blue-green photons on the one hand, and are able to quench (bacterio)chlorophyll triplet states with high efficiency on the other[1]. (Bacterio)chlorophyll triplets arise through intersystem crossing with a relatively high yield, and may sensitise reactive species of oxygen by excitation energy transfer (e.g., triplet-singlet transfer to the highly-oxidising singlet oxygen). Photoprotection against triplet states is observed in both oxygenic and anoxygenic photosynthesis, and it is strikingly more efficient in oxygenic organisms, where the probability of transfer to oxygen is higher [2]. In light-harvesting (LH) complexes from purple bacteria, the triplet-triplet transfer between BChl and carotenoid was extensively studied in the 1990s, and shown to occur in the nanosecond range [3]. However, more recent studies have provided a more complex picture of the triplet management in these complexes, and, despite their functional importance, the different molecular mechanisms involving these key excited states of carotenoids and (bacterio)chlorophylls remain unclear.

The reaction centre (RC) and LH pigment-protein complexes of purple bacteria are among the most extensively studied systems in biology. Since the development of the first isolation protocols and associated spectroscopic studies in the 1970s [4-6], they have served as benchmark models for investigating excitation energy transfer. The LH family comprises two main types of proteins - LH1, tightly associated with the reaction centres, and LH2 which act as peripheral antennae, transferring the excitation energy to the RC *via* the LH1 complexes. Both LH1 and LH2 are doughnut-shaped pigment-proteins, composed of a multimeric association of a minimal repeat unit itself comprising two small, trans-membrane-spanning apoproteins (α- and β-) which bind the bacteriochlorophyll cofactors non-covalently [7]. In all LHs, a concentric ring of strongly-coupled BChls is responsible for the lower energy transition of the protein [8-9] and are denoted B850 and B875 in LH2 and LH1, respectively [10]. In LH2, following carotenoid excitation, the resulting $S_2$ excited state evolves on sub-picosecond timescales through competing pathways, involving internal conversion into the dark states of the molecule manifold and energy transfer to nearby B850 BChls. In complexes hosting shorter carotenoids, dark states such as $S_1$ can also contribute to B850 population via measurable $S_1 \rightarrow$B850 energy transfer [11-13]. On longer timescales, carotenoid triplet signals are clearly observed in purple bacterial light-harvesting complexes containing carotenoids of different conjugation lengths (N), including neurosporene (N = 9) and spheroidene (N = 10) in *Cereibacter* (formerly *Rhodobacter*) *sphaeroides* 2.4.1 LH2 [14], rhodopin glucoside (RG) (N = 11) in *Rhodoblastus* (*Rbl.*) *acidophilus* 10050 LH2, and spirilloxanthin (N = 13) in *Rhodospirillum rubrum* S1 LH1 [3, 15]. It was proposed that carotenoid triplets are populated *via* triplet–triplet energy transfer from bacteriochlorophyll triplets formed through intersystem crossing [3]. In *Rbl. acidophilus* LH2 the BChl-*a*-to-carotenoid triplet–triplet transfer has been measured between 0.45 ns [16] and 20 ns [3]. However, more recent transient absorption studies on LH1 from *Rhodospirillum rubrum* S1, containing spirilloxanthin (N = 13), reported an S* state and carotenoid triplet absorption appearing within a few picoseconds after carotenoid excitation. This S* was interpreted as an intermediate entangled triplet $^1$(TT) seeding ultrafast triplet formation via singlet fission [17]. A subsequent study on LH2 from *Cereibacter sphaeroides* (spheroidene, N = 10) similarly concluded that S* is populated directly from $S_2$, contributes to energy transfer toward BChl, and precedes a long-lived carotenoid triplet spectrum on the picosecond timescale, again consistent with triplet formation through a singlet-fission branch [18]. This view aligns with studies conducted in polyenes and carotenoids showing that an underlying spin-entangled triplet-pair character provides the mechanistic basis for producing longer-lived, separated triplet pairs under specific conditions [19], and the observation in carotenoid aggregated systems that S* is the precursor to long-lived carotenoid triplets [20]. However, the argument that the long-living triplet is favoured by protein-imposed twisting of the carotenoid backbone is somewhat contradicted by studies on the Orange Carotenoid Protein - where no long-living triplets are observed in the twisted carotenoid form [21]. Recently, singlet fission has been proposed to mediate carotenoid-to-bacteriochlorophyll *a* energy transfer in purple bacterial RC–LH1 complexes, through a heterofission mechanism in which $S_2$ excitation generates a triplet pair partitioned between a carotenoid and a neighbouring BChl *a* [22]. This assignment rests largely on magnetic field effects and on weak near-infrared transient absorption features, without a mode-selective probe of the carotenoid excited state itself. Whether such a process operates in LH2, and if the reported signatures require a heterofission assignment rather than a carotenoid-localised triplet pair, remain open questions. The origin of these conflicting interpretations lies in methodological limitations of using femtosecond transient absorption spectroscopy (TA) as the main probing technique. In LH complexes, multiple pigments contribute overlapping ground-state bleaching (GSB) and excited-state absorption (ESA) features. As a result, transient-absorption datasets often contain overlapping signals and rely strongly on global or target analysis models, particularly in the congested carotenoid region, where dark-state signatures are broad and not uniquely distinguishable. Femtosecond stimulated resonance Raman spectroscopy (FSRRS), which isolates specific excited states and amplifies the Raman gain through resonance enhancement [23-25], can circumvent the limitations of electronic techniques.

Recently, we have obtained direct fingerprints of different excited states in isolated carotenoids [26], providing a spectral library of the vibrational modes of the excited states which can be used for precise mapping of the processes operating in photosynthetic proteins. In the present work, we apply FSRRS to the well-characterized LH2 antenna from the prototypical purple bacterium *Rbl. acidophilus* [27]. Its ring structure (Figure 1a), determined by X-ray crystallography over 30 years ago, is composed of nine repeat units, each comprising a pair of α- and β-apoproteins that noncovalently bind three molecules of BChl *a* and one carotenoid molecule, rhodopin glucoside (RG) [8, 28-29]. Two of these BChls participate in the strongly coupled ring of 18 BChls responsible for the 850 nm absorption of the protein, and the third is part of the weakly-interacting ring of 9 BChls at the origin of absorption at 800 nm. The RG carotenoids establish *van-der-Waals* contacts with both BChl groups [8], and their $S_1$ energy is comparatively low, reported as circa 12,550 $cm^{-1}$, giving a correspondingly weak $S_1$-mediated energy transfer to BChl and positioning $S_2$ as the dominant donor channel. Because this $S_1$ energy lies above the BChl *a* Qy transition (circa 11,700 $cm^{-1}$ at 850 nm), BChl *a* cannot transfer energy back into the carotenoid singlet manifold via the reverse, Qy-to-$S_1$ pathway. This asymmetry facilitates the study of triplet populations [30].

## Results

### *Electronic characterisation of ground and excited states*

The molecular structure of LH2 complex from *Rbl. acidophilus* 10050 is shown in Figure 1a (PDB 1NKZ)[28]. Three separated pigment pools contribute to the room-temperature absorption spectrum - RG, B800, and B850 - allowing selective excitation of either the carotenoid or the BChl *a* subsystem (Figure 1b). Femtosecond TA following carotenoid excitation at 485 nm shows an ESA maximum at 578 nm with a shoulder near 550 nm (Figure 1c and supporting information Fig. S1). The 578 nm band decays completely within a few picoseconds, whereas the 550 nm shoulder decays only partially over the same period and then grows slightly, persisting beyond the 7 ns experimental window. This long-lived feature has previously been attributed to a carotenoid triplet state [16]. In the near-IR, an ESA at 830 nm and a ground-state bleach at 855 nm correspond to BChl *a* excitation, decaying over a few nanoseconds. Excitation of B850 at 875 nm produces no detectable carotenoid signal at early time delays, consistent with the absence of efficient B850-to-RG energy transfer (Figure 1d and Supporting Information, Fig. S2). The near-IR ESA and bleach at 830 and 855 nm, respectively, are again present, together with very weak BChl *a*-associated features between 500 and 650 nm, with a wide peak at 540 nm. Most of these features decay synchronously with the near-IR dynamics; however, a wide component with maximum at 550 nm persists beyond the experimental window and even grows slightly, mirroring the long-lived feature seen upon carotenoid excitation. The overall carotenoid-to-BChl *a* energy-transfer efficiency (about 50% *via* $S_2$ and 5% *via* $S_1$) is consistent with previous reports [16, 30-31]. The long-lived 550 nm feature is compatible with two interpretations. It may report on a carotenoid triplet seeded by singlet fission from its own S* state, as proposed for LH1 and LH2 complexes containing shorter carotenoids [17-18]. Alternatively, it may be assigned to two spectrally overlapping but mechanistically independent species, favouring the T-T energy-transfer interpretation [16]. It is worth noting that, at early times after BChl *a* excitation, we do not detect any spectral feature attributable to a carotenoid triplet, which argues against the involvement of a heterofission process. However, due to the width of features in the 530-560 nm region, TA alone cannot discriminate between these possibilities, which motivates the state-selective FSRRS experiments reported here.

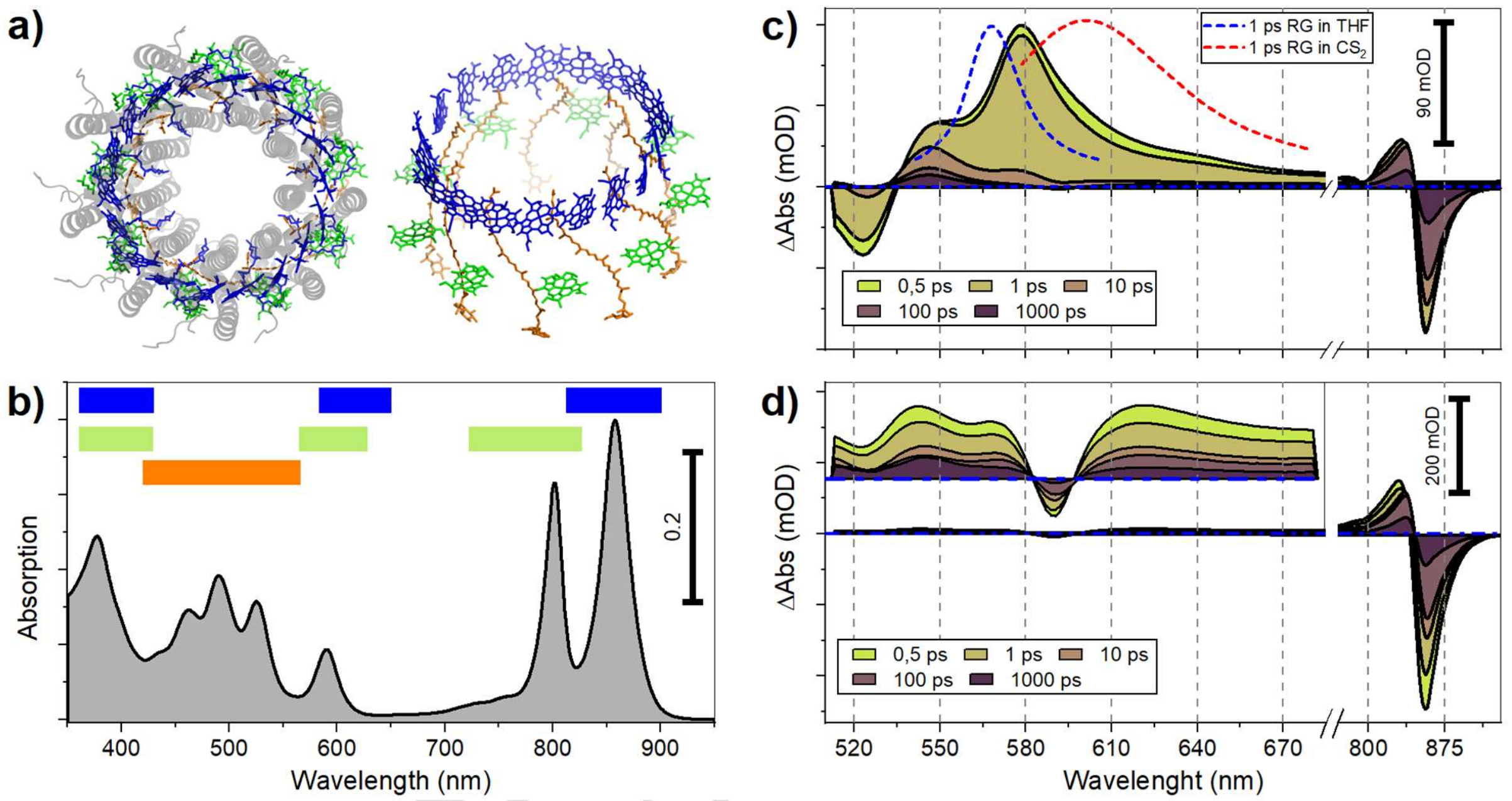

**Figure 1 |** Crystal structure, absorption, and ultrafast spectroscopy of the LH2 complex from *Rbl. acidophilus* 10050. (a) X-ray crystal structure (PDB 1NKZ) showing bacteriochlorophyll a in the B800 (green) and B850 (blue) rings and the carotenoid RG (orange). (b) Room-temperature absorption spectrum; coloured bars indicate the principal absorption regions of B800 (green), B850 (blue), and RG (orange). (c) Femtosecond TA spectra upon 485 nm carotenoid excitation at the indicated pump-probe delays. The dashed spectra are RG in solution at 1 ps under the same actinic pump: blue, RG in THF (ESA maximum at 568 nm); red, RG in $CS_2$ (ESA maximum at 600 nm). Full delay series in Figs. S3 (THF) and S4 ($CS_2$). (d) Femtosecond TA spectra upon 875 nm B850 excitation at the indicated delays; the horizontal blue dashed line marks zero differential optical density (ΔOD). Inset: expanded view, 510–680 nm.

To establish the spectral signatures and lifetimes of the RG dark states, we measured the TA of RG in tetrahydrofuran (THF) with the same 485 nm actinic pump (Supporting Information, Fig. S3). The 1 ps gated spectrum (blue dashed line, Figure 1c) shows a single ESA maximum at 568 nm, red-shifted by 10 nm from the value reported in acetone[31]. Unlike RG in LH2, the THF spectrum shows no pronounced shoulder on the blue side of the ESA. Global sequential analysis of the RG-in-THF dataset is satisfactorily described by four components (100 fs, 0.4 ps, 3.7 ps, 7.3 ps; full delay series, 2D map, fit and residuals in Figure S3). By comparison with lycopene ($N_{eff}$= 11), which follows the same kinetic scheme [26], we assign the 100 fs component to $S_2$ decay (convolved with the instrument response function and treated as an upper limit), 0.4 ps to hot-$S_1$, 3.7 ps to $S_1$, and 7.3 ps to S*. The same measurement was carried out for RG in $CS_2$ (Supporting Information, Fig. S4), the most polarisable solvent tested. The 1 ps gated spectrum (red dashed line, Figure 1c) shows a single ESA maximum at 600 nm, and the band is considerably broader than in THF (and LH2). Global sequential analysis of the RG-in-$CS_2$ dataset is described very well by four components (100 fs, 0.4 ps, 4.1 ps, 13.6 ps; full delay series, 2D map, fit and residuals in Figure S4), assigned as above to $S_2$, hot-$S_1$, $S_1$ and S*. The $S_1$ lifetime is essentially unchanged between the two solvents (3.7 and 4.1 ps), although the ESA envelope widens. Subsequently, the S* lifetime increases from 7.3 ps in THF to 13.6 ps in $CS_2$, consistent with values reported previously [32-33]. Note that RG in the more polarisable solvent $CS_2$ shows an $S^*/^1(TT)$ lifetime that increases to 13.6-18 ps, indicating that solvent polarisability alone already lengthens this state in isotropic media, an effect widely described in the literature [31-33].

***Vibrational characterisation of excited states***

The TA data establish the approximate spectral positions and lifetimes of the excited species, both BChl *a* and carotenoid dark states, but TA alone cannot resolve the origin of the transient species that overlap in the congested 500-560 nm region. We therefore recorded FSRRS datasets with the actinic pump (AP) fixed at 485 nm, exciting RG to $S_2$, and the Raman pump (RP) tuned across the resonance conditions of the different carotenoid dark excited-state species. We have previously used this approach to demonstrate state selectivity for a series of isolated carotenoids in solution [26]; here we apply it for the first time to a carotenoid bound to its native protein host.

Resonance Raman spectra of carotenoids in the ground state (Supporting Information, Fig. S5) display four characteristic band groups, $\nu_1$ to $\nu_4$. The $\nu_1$ band, above 1500 $cm^{-1}$, arises from C=C stretching of the conjugated backbone and is sensitive to conjugation length and molecular configuration. The $\nu_2$ envelope, near 1160 $cm^{-1}$, corresponds to C-C stretching coupled to in-plane C-H bending, and is diagnostic of *cis*/*trans* isomerisation. The $\nu_3$ band, near 1000 $cm^{-1}$, arises from in-plane rocking of the methyl groups attached to the conjugated chain, coupled to adjacent C-H bending, and is diagnostic of end-ring configuration [34]. In the excited-state manifold revealed by FSRRS, the $\nu_1$ region (1400 to 1800 $cm^{-1}$) is the most affected, resolving into four sub-bands, $\nu_{1a}$ to $\nu_{1d}$, each reporting on a distinct dark state: $\nu_{1a}$ on $S_1$, $\nu_{1b}$ on hot-$S_1$, $\nu_{1c}$ on an ICT-like state, and $\nu_{1d}$ on $S^*/^1(TT)$ [26].

*Rhodopin glucoside photoexcitation in solvents*

In order to characterize and assign the excited state phenomena occurring in the protein, we first need to observe the vibrational features associated with isolated RG. We have measured the vibrational modes of RG excited states upon 485-nm-excitation, for the carotenoid in THF (RP 530, 540 nm) and $CS_2$ (RP 550, 600 nm) (Supporting Information, Figs. S6, S7, S9 and S10). Figure 2a shows the gated spectra for RG in THF, revealing the maxima of the more prominent vibrational modes at 1142, 1215, 1252, 1471, 1540, and 1780 cm$^{-1}$. We can observe how the features close to 1480 cm$^{-1}$ - associated with S*/$^{1}$(TT) - decay on a timescale of a few ps, slightly longer than the $S_1$-associated signals. As seen in Figure 2b, the gated spectra in $CS_2$ at RP 550 nm are much less defined, revealing only a small feature *circa* 1780 cm$^{-1}$, and two features at 1143 and 1489 cm$^{-1}$. The $S_1$-associated feature at 1780 cm$^{-1}$ is much clearer upon resonance excitation at RP 600 nm (Supporting Information, Fig. S10). Comparing the kinetic traces, it is clear that whereas the $S_1$-associated mode decays at the same rate for both solvents, the S*/$^{1}$(TT) feature rises more slowly and is considerably longer-lived in $CS_2$. Note that a similar dependence on solvent polarisability has recently been reported for lutein, with a two-fold lengthening of the S*/$^{1}$(TT)-associated lifetimes to 29 ps in $CS_2$, comparable to the effect observed here [35].

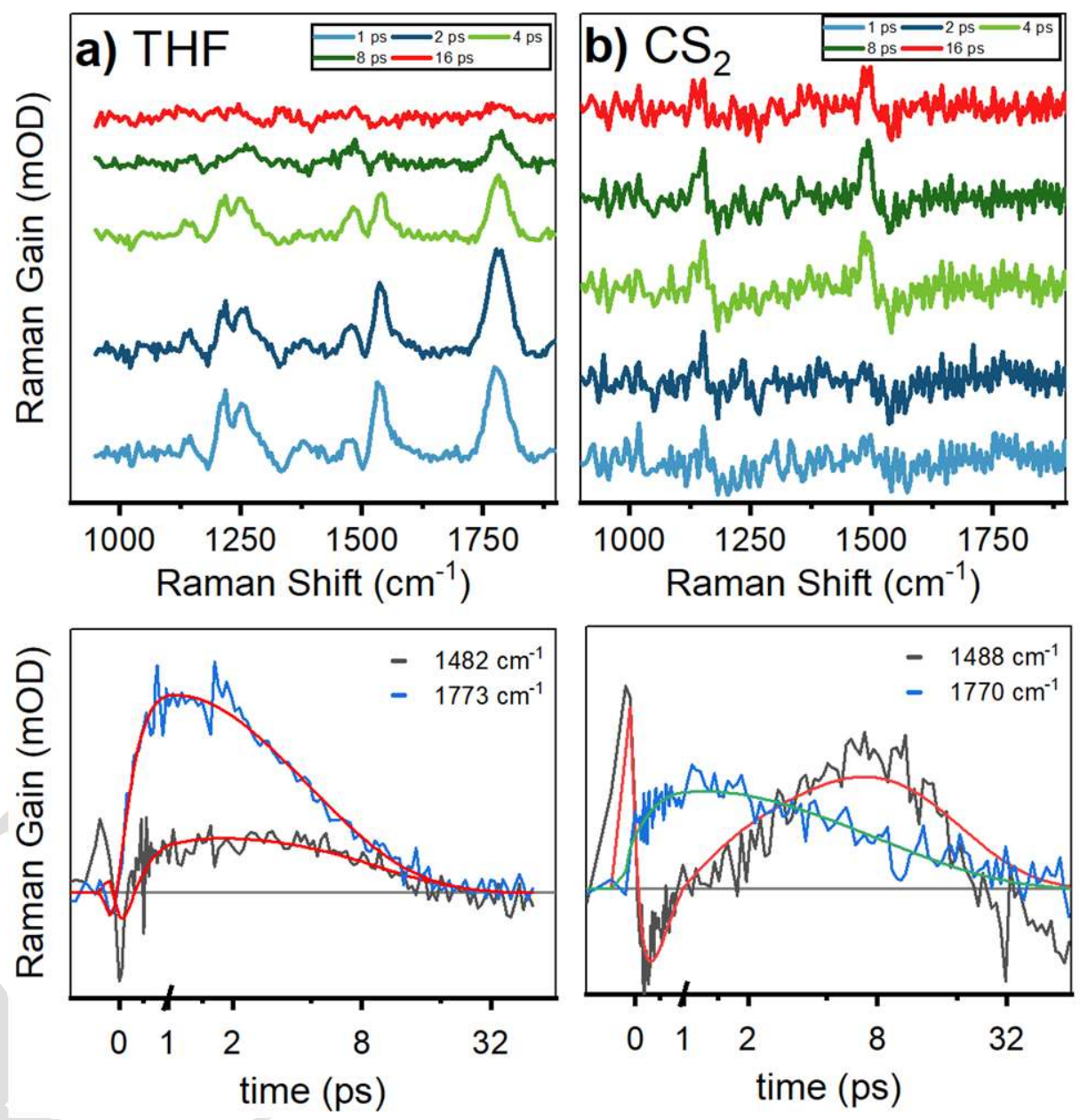

**Figure 2 |** Resonance-selective FSRRS of rhodopin glucoside in THF (a) and in $CS_2$ (b). Actinic pump 485 nm in both cases. The Raman pump was set to 540 nm (THF) and 550 nm ($CS_2$), matching the excited-state absorption of the dark-state manifold in each solvent. Upper panels: time-gated spectra at selected delays, colour coded from early (blue) to late (red) times. Lower panels: kinetic traces taken at the maxima of the v1a and v1d bands.

*Rhodopin glucoside photoexcitation in LH2*

We obtained maps for isolated LH2 at actinic pump 485 nm with the RP set from 520 to 600 nm (Supporting Information, Figs. S11-S16). Figure 3 presents three representative matrices, at RP 530, 560, and 600 nm, showing for each the frequency-time 2D map, time-gated spectra, and kinetic traces. Across all RP wavelengths, LH2-bound RG exhibits the same four-$\nu_1$ mode pattern ($\nu_{1a-d}$) reported for isolated carotenoids [26, 35], and these $\nu_1$ modes remain the most readily-distinguishable markers of the underlying dark-state population.

At RP 530 nm, resonant with $S^*/^1(TT)$ excited-state absorption (Figure 3a), $\nu_{1d}$ decays mono-exponentially. No residual signal grows in at later delays, confirming that $S^*/^1(TT)$ decays directly to the ground state without producing a secondary long-lived species. At early delays, the associated $\nu_2$ mode appears at 1138 $cm^{-1}$ and downshifts to 1130 $cm^{-1}$, while $\nu_{1d}$ itself shifts from 1484 to 1478 $cm^{-1}$ over the same period, from a few picoseconds to nanosecond delays.

At RP 560 nm, resonant with $S_1$ stimulated absorption as well as with $S^*/^1(TT)$ (Figure 3b), the spectra contain all the modes ($\nu_{1b}$ only weakly observed; clearer at RP 600 nm - see below): $\nu_{1a}$ at 1764 $cm^{-1}$, $\nu_{1c}$ at 1541 $cm^{-1}$, and $\nu_{1d}$ at 1482 $cm^{-1}$, accompanied by $\nu_2$ at 1134, 1206, and 1271 $cm^{-1}$. On the nanosecond scale the modes show an apparent downshift, less pronounced than at RP 530 nm, within the limit of our resolution. Knowing that the vibrational modes arise from the same excited states, the difference from the RP 530 nm downshift can be explained by the interplay of resonances from two states with very similar Raman modes.

At RP 600 nm the spectra are dominated by a broad feature in the 1700-1750 $cm^{-1}$ region ($\nu_{1b}$) peaking at 1731 $cm^{-1}$, followed by $\nu_{1a}$ at 1765, $\nu_{1c}$ at 1545, and $\nu_{1d}$ at 1484 $cm^{-1}$. This matrix shows no clear nanosecond signature, indicating that at this RP, the state responsible for the long-lived signal absorbs very weakly at most. The $S^*/^1(TT)$ vibrational markers at 1484 and 1134 $cm^{-1}$ are nonetheless present, indicating that, despite the similarity in their vibrational modes, $S^*/^1(TT)$ and the long-lived species are two different states.

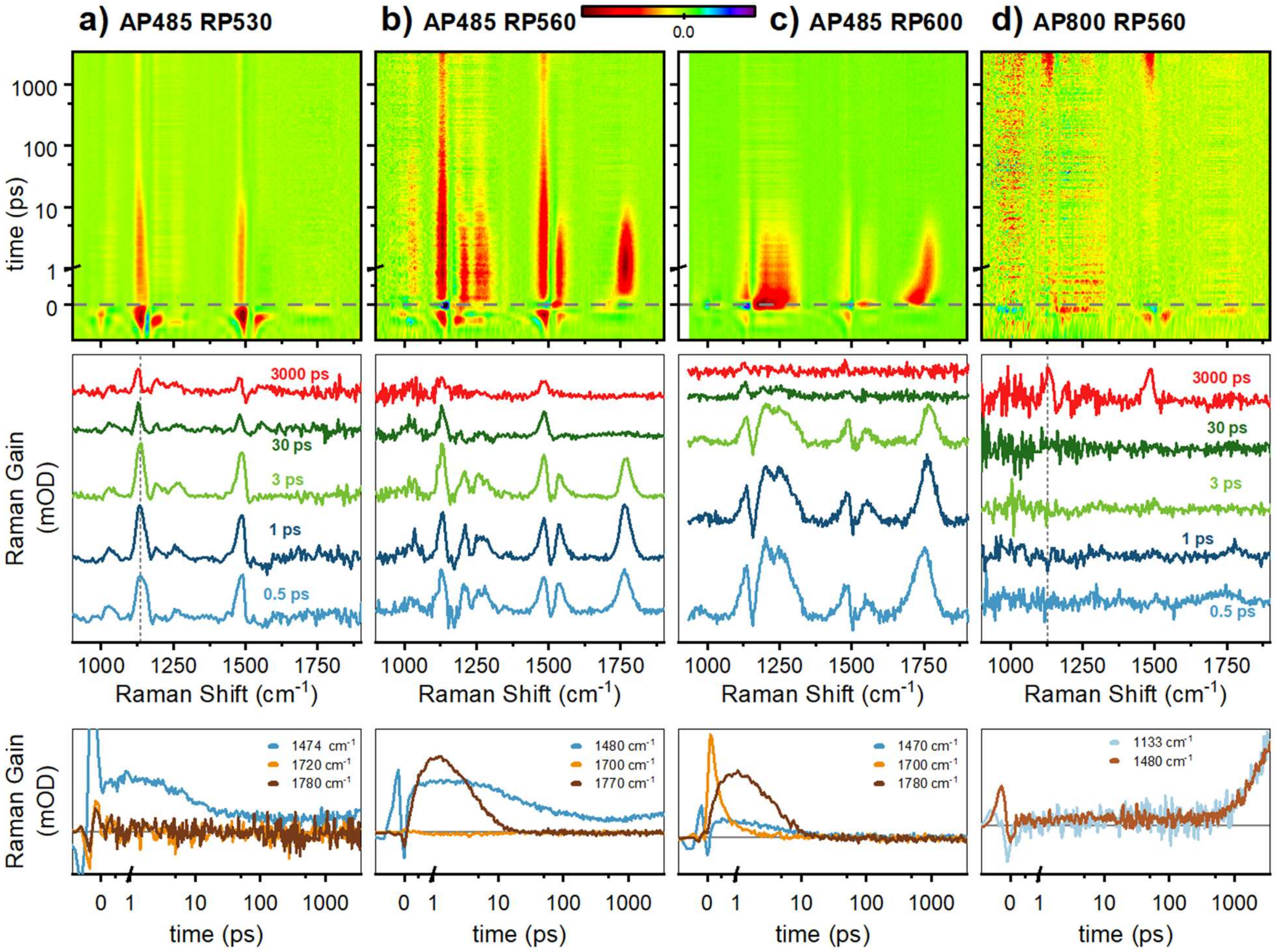

**Figure 3 |** Resonance-selective FSRRS of carotenoid dark states in LH2 under carotenoid and BChl *a* excitation. Actinic pump: 485 nm for (a-c). Raman pump tuned to (a) 530 nm; (b) 560 nm; (c) 600 nm; and (d) actinic pump 800 nm (BChl *a*), RP 560 nm. For each: (top) frequency-time colourmap of Raman gain as a function of Raman shift ($cm^{-1}$) and pump-probe delay; (middle) time-gated spectra at selected delays, with the y-scale (mOD) inside each panel; (bottom) kinetic traces of the assigned $\nu_1$ modes.

*BChl a photoexcitation in LH2*

The $S_1$ energy of RG lies above the B850 $Q_y$ transition, so carotenoid dark states cannot be populated by singlet energy transfer from BChl *a*. Exciting the BChl *a* $Q_y$ band at 800 nm therefore isolates any triplet-formation channel that does not proceed through the carotenoid singlet manifold (Supporting Information, Fig. S17). Additionally, Raman pumping in the 530 – 600 nm region ensures the detection of carotenoid species only. At early delays we detect no $S_1$, hot-$S_1$, or $S^*/^1(TT)$ signatures in the $\nu_1$ region (Figure 3d), confirming the absence of a carotenoid singlet precursor under these conditions. Instead, two carotenoid-associated vibrational modes at 1130 and 1484 $cm^{-1}$ rise over the hundreds-of-picoseconds timescale. These modes have previously been associated with the carotenoid triplet on microsecond timescales [36-37] and by power-induced Raman [2]. In the absence of any signal from the carotenoid singlet dark states, their rise indicates population of the carotenoid $T_1$ state by triplet-triplet transfer from BChl *a* following intersystem crossing. It is worth noting here that we do not observe any signal associated with carotenoid $^{1}$(TT) or triplet at early times, which should be clearly distinguishable in the case of heterofission.

***Multi-Matrix Global Analysis***

The resonance effect can take us beyond merely analysing each matrix separately. Each FSRRS matrix is bilinear: at any given delay, the measured spectrum is a sum of a few components, and each component is a fixed vibrational spectrum multiplied by a population that changes with time. The time evolution of the different populations are related *via* (Equation 1)

$$\frac{d\mathbf{c}(t)}{dt} = \boldsymbol{K}\,\mathbf{c}(t), \qquad \mathbf{c}(0^{+}) = \mathbf{j} \qquad (1)$$

The vector $\mathbf{c}(t)$ holds the populations, $\mathbf{j}$ is the initial population set by the excitation, and $\boldsymbol{K}$ is the transfer matrix. The dataset therefore factorises into a concentration matrix $\mathbf{c}^{\mathrm{T}} = \mathbf{C}$, which holds the time-dependent populations, and a spectral matrix $\mathbf{S}$, which holds the fixed component spectra (Equation 2) [38].

$$\boldsymbol{\Psi}(t,\tilde{\nu}) = \sum_{l=1}^{n_c} c_l\,(t)\,s_l(\tilde{\nu}) = \boldsymbol{C}\,\boldsymbol{S} \qquad (2)$$

In our particular case, for the sake of clarity we assume that populations follow a first-order compartmental scheme, in which each species decays into the next with a characteristic lifetime, while the spectra enter the model linearly. We used this split through variable projection: only the non-linear parameters (the lifetimes, time-zero, and the width of the instrument response function) are searched, and the spectra are then reconstructed analytically at each step rather than fitted [38]. The full derivation is given in the Supporting Information. The matrices recorded under different resonance conditions share a single kinetic model. The same species are probed whatever the Raman-pump wavelength, so one set of lifetimes $\boldsymbol{\theta}$ describes every matrix of a given sample; changing the Raman-pump wavelength changes only how strongly each species contributes, not when it appears or decays. We used this constraint in two ways. Firstly, the vibrational fingerprint of a species is fixed by its normal modes and does not change from one matrix to the next. Its spectrum in each matrix is therefore a single shared band shape scaled by a per-matrix amplitude, with the first amplitude fixed to set the overall scale (Equation 3).

$$\boldsymbol{S}_{\boldsymbol{i,c}}(\tilde{\nu}) = \alpha_{i,c}\,\boldsymbol{S}_{\boldsymbol{c}}(\tilde{\nu}), \qquad \alpha_{1,c} = 1 \qquad (3)$$

Secondly, we fitted all matrices at once by minimising a single objective summed over the whole set (Equation 4), with the shared band shapes and the amplitudes as the linear unknowns and only $\boldsymbol{\theta}$ searched non-linearly.

$$\chi^2 = \sum_{i=1}^{M} \|\boldsymbol{\Psi}_i - \boldsymbol{C}_i(\boldsymbol{\theta})\, \boldsymbol{S_i}\|^2 \qquad (4)$$

Linking the matrices in this way has two benefits. A species that is weak or short-lived in one matrix can be extracted using information from the matrices in which it is better resolved. It also constrains the degrees of freedom for a species to take different band shapes in each matrix, reducing the likelihood that spectral noise can imitate a real spectral change. We applied the link only over the Raman-shift ranges where the fingerprint is invariant, and left the windows that are genuinely matrix-dependent free to differ.

We applied the multi-matrix global analysis with a sequential model to the entire dataset associated to each sample, to obtain the positions of the associated modes and their lifetimes (Figure 4 and Supporting Information, Fig. S18). Individual fits, residuals, EADS, and kinetic traces are provided in the Supporting Information, Figs. S6-S17. The lifetimes were first obtained from vibrational modes that are isolated from, or independent of, the other species. The formation lifetime of the long-lived species was obtained from the BChl-excitation dataset (AP 800 nm, RP 560 nm), giving 2100 ps. The $S_1$ lifetime was obtained from an individual fit of the $\nu_{1a}$ mode near 1780 $cm^{-1}$, which is completely isolated at AP485, RP 560 nm. At AP485, RP 530 nm, where $S^*/^1(TT)$ is largely dominant, the modes near 1480 $cm^{-1}$ gave an $S^*/^1(TT)$ lifetime of 60 ps. These values were used as initial parameters and were allowed to evolve. Figure 4 shows the selected EADS extracted for RG in THF, and for LH2 under carotenoid excitation; for clarity only the $S_1$, $S^*/^1(TT)$, and long-lived components are shown (all EADS and fittings for all the discussed samples are in the Supporting Information).

For RG in THF, the $S_1$-associated $\nu_{1a}$ mode appears at 1778 $cm^{-1}$ (3.7 ps decay); the $^1(TT)$-associated spectrum exhibits its $\nu_2$ mode at 1140 $cm^{-1}$ and $\nu_{1d}$ at 1492 $cm^{-1}$, together with a band at 1790 $cm^{-1}$, which we designate $\nu_{1a}$*. In LH2, the $S_1$ $\nu_{1a}$ band appears at 1772 $cm^{-1}$ and again decays in 3.7 ps. The $S^*/^1(TT)$-associated modes appear at 1130 $cm^{-1}$ ($\nu_2$) and 1482 $cm^{-1}$ ($\nu_{1d}$), accompanied by a much weaker $\nu_{1a^*}$ mode at 1782 $cm^{-1}$. This $\nu_{1a^*}$ band, upshifted by 10-12 $cm^{-1}$ from $\nu_{1a}$, appears in all matrices, whereas it is absent in reported spectra of separated, non-entangled triplets [2, 37] (formation at 2100 ps, and the non-decaying component). Relative to solution, the $S_1$ lifetime is unchanged while the $S^*/^1(TT)$ lifetime increases from 7.3 ps to about 60 ps; a 2100 ps component and a non-decaying component are also recovered, the latter previously attributed to the carotenoid triplet formed by triplet-triplet energy transfer from BChl-T.

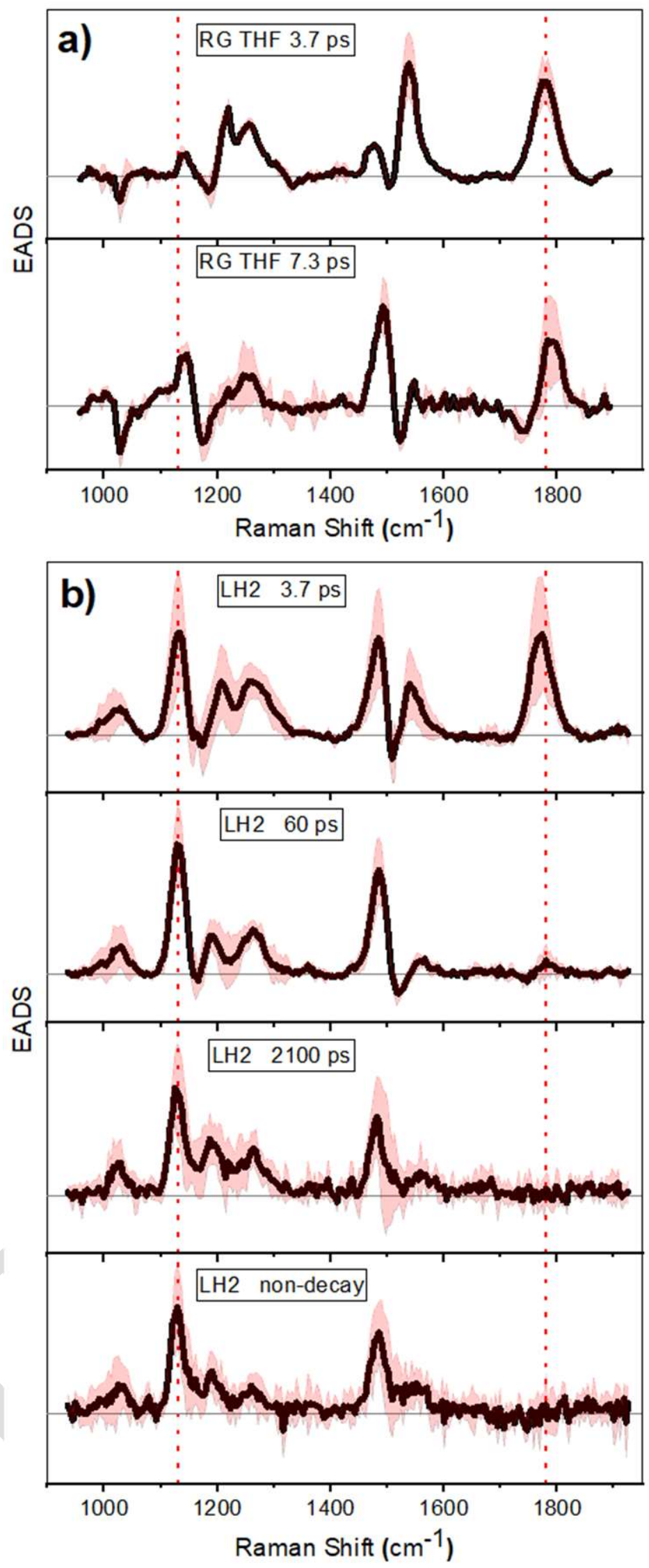

**Figure 4 |** EADS obtained from multi-matrix global fitting using sequential analysis. **a)** representative EADS for RG in THF. **b:** representative EADS for LH2, showing the associated lifetimes for the $S_1$, $S^*/^1(TT)$, and long-lived components; the standard deviation is shown as the shaded red band. The dashed lines at 1130 and 1785 $cm^{-1}$ act as visual aids to emphasise the discussed band-shifts. All spectra and fittings are provided in the Supporting Information, Fig. S18.

## Discussion

### *The entangled pair and the $\nu_{1a^*}$ mode*

In solution, RG follows the excited-state cascade expected for its conjugation length (N = 11), the $S_2$, hot-$S_1$, $S_1$, $S^*/^1(TT)$ sequence reported for lycopene [26], with the same $\nu_{1a}$ to $\nu_{1d}$ markers. Beyond the $\nu_{1d}$ marker, the $S^*/^1(TT)$ spectrum carries a further high-frequency mode, upshifted by 10 to 12 $cm^{-1}$ from $\nu_{1a}$ (ca. 1790 $cm^{-1}$), designated $\nu_{1a^*}$ here. A comparable feature has been reported for spirilloxanthin [39] but was not clearly resolved in other carotenoids studied by FSRS [26, 40]. In isolated carotenoids the short S*/$^1$(TT) lifetime keeps $\nu_{1a^*}$ poorly separated from $\nu_{1a}$, such that it may not appear in the global model. In LH2 the S*/$^1$(TT) lifetime is about 60 ps, long enough to isolate $\nu_{1a^*}$ cleanly, and the same 10 to 12 $cm^{-1}$ upshift observed for RG in THF is then recovered in LH2. We therefore assign $\nu_{1a^*}$ to a genuine mode of the S*/$^1$(TT) state, and thus a vibrational signature reporting on the structure of the $^1$(TT) state. In the isolated carotenoid, its proximity to $\nu_{1a}$ and the short $S_1$ lifetime meant it was easily dismissed as leakage from $S_1$ in the sequential fit. In LH2 this reading no longer holds: the S*/$^1$(TT) lifetime is more than an order of magnitude longer than $S_1$, $\nu_{1a^*}$ appears in every matrix, and it is absent from the spectrum of isolated triplets [2, 37]. A high-frequency mode upshifted from $\nu_{1a}$ is consistent with triplet-pair character that remains coupled to the ground state, as described for the S1 mode, and its position may provide a route to estimating the energy of the $^1$(TT) state [41]. As another possibility within the framework that relates the proposed dark states to correlated triplet-pair configurations, the $\nu_{1a^*}$ band may reflect an Ag-like component of the $^1$(TT) wavefunction, as expected from calculations by Tavan and Schulten [42]. Signals in the same spectral region have previously been reported by Kloz et al. [39] and, more recently, by Chrupkova *et al.*[43], supporting the assignment of Ag-type vibrational character within the triplet manifold. In this view, the observation of $\nu_{1a^*}$ in the triplet pair – previously assigned as [$^3$Bu ⊗ $^3$Bu] – is consistent with a strongly coupled Bu/Ag manifold, where geometry modulates the mixing between Bu- and Ag-like configurations rather than defining fully-distinct electronic states. Although this last point is tentative and will require dedicated experiments, it is consistent with reported mixing in the singlet manifold [26, 44] . The entangled-pair fingerprint, $\nu_{1a^*}$ included, is carried by a carotenoid-localised state and is distinct from the modes of the bare carotenoid triplet. The lifetime of the entangled pair is the one property that changes between isolated and protein-bound carotenoid. In solution it is about 7.3 ps (RG in THF); the most polarisable solvent tested, $CS_2$, lengthens it to the 13.6 - 18 ps range [32-33]. Thus, environment polarisability alone can account for about a third of the protein-associated slowing. The remaining lengthening, to about 60 ps, requires a further contribution, which we attribute to the restricted conformational freedom imposed by the binding pocket. For an entangled pair, twisting the backbone can decouple the two triplet contributions and lengthen the state lifetime. We know that RG is situated in a highly-polarisable environment [34, 45-46] and that its protein binding pocket imposes a twisted conformation [8, 47]. Polarisability provides a first, modest stabilisation of the triplet correlated pair, if we consider this as an increase in the lifetime, whereas the binding pocket shifts the conformational ensemble towards geometries that further stabilise the pair, accounting for the remaining lifetime extension.

***Decay of the entangled pair in LH2***

In LH2 we can isolate the entangled pair from the independent triplet despite their overlapping features. The main bands at 1140 cm$^{-1}$ and $\nu_{1d}$ at 1492 cm$^{-1}$ lie at the limit of resolution between $^1$(TT) and the independent triplet, but the selective resonance approach separates them. The pair forms from $S_2$ in under 150 fs and decays to the ground state with a lifetime of about 60 ps. At RP 530 nm no residual or growing feature appears beyond the S*/$^1$(TT) decay, which indicates direct return to the ground state. Stabilization of intramolecular $^1$(TT), through protein-imposed torsion, increases the lifetime by an order of magnitude, but does not result in a long-living T species as proposed elsewhere, based on TA measurements [17-18]. As already discussed, the Orange Carotenoid Protein shows no efficient long-lived triplet generation from $^1$(TT), despite the presence of a twisted carotenoid under comparable conditions [21]. Our results are fully consistent with this observation: twisting lengthens the entangled pair without generating separated triplets. This observation is also supported by the distances retrieved from the crystal structure. In order to dissociate the $^1$(TT) into two independent triplets residing on different carotenoids, the nearest-neighbour centre-to-centre distance should be about 4.5 Å or less, beyond which the coupling is too weak, both for carotenoids and for other molecules generally reported [48-51]. In LH2 the closest carotenoid-carotenoid centre-to-centre separation is about 18.1 Å from structural analysis (Figure 5). Even the edge-to-edge distance is far beyond the 4.5 Å limit. The same separation rules out both the formation of two independent triplets on adjacent carotenoids, and an entangled pair shared between two carotenoids [52-53]. The long-lived triplet cannot arise this way.

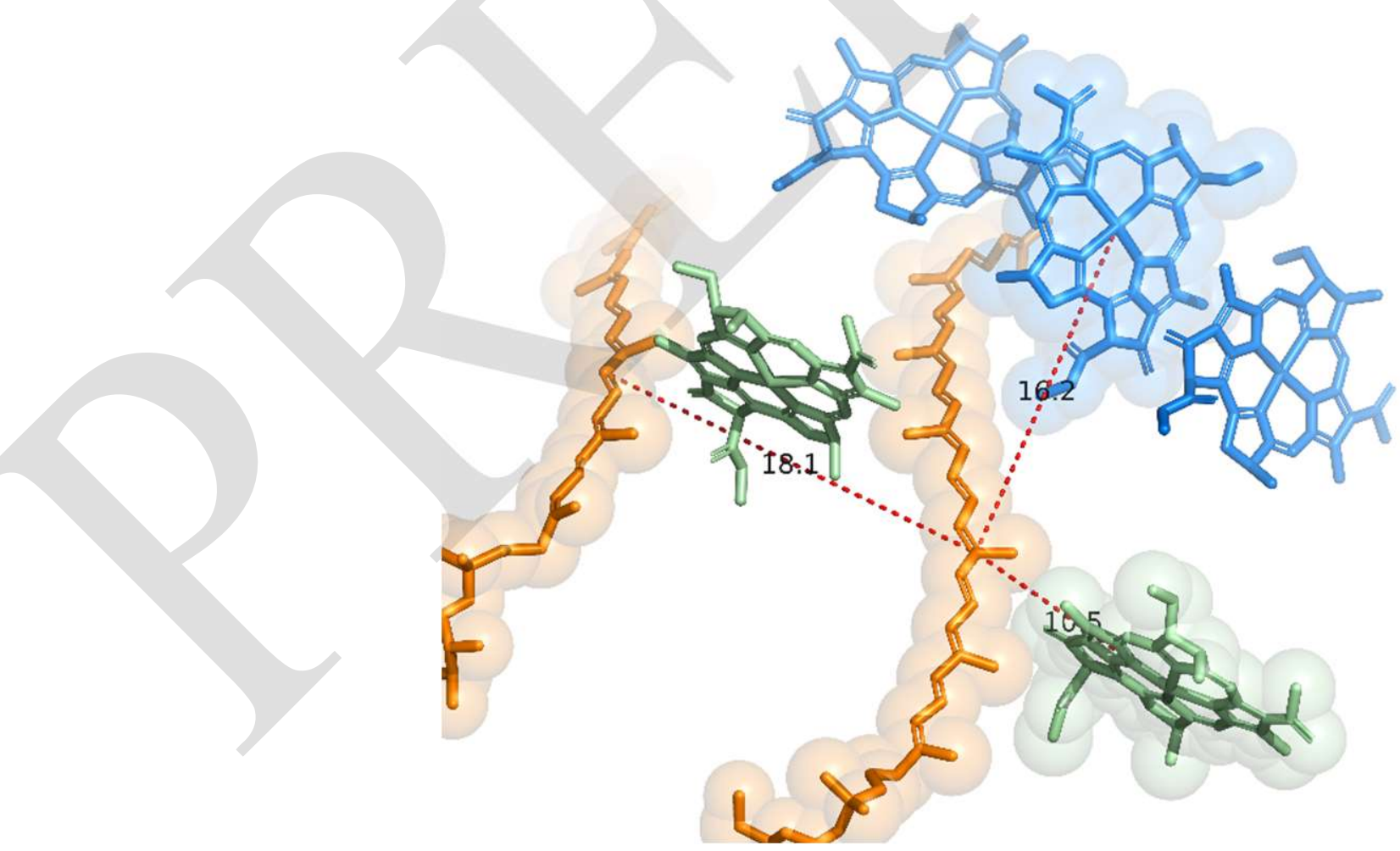

**Figure 5 |** Detail of the crystal structure of the LH2 complex from Rbl. acidophilus 10050 (PDB 1NKZ). It shows the position and the distances from core to core of bacteriochlorophyll a in the B800 (green) and B850 (blue) rings and the carotenoid RG (orange). The red dashed lines show the distance centre-to-centre between RG and the nearest RG (18.1 Å), nearest B800 (10.5 Å), and the nearest B850 (16.2 Å).

### *High-precision detection of triplet-triplet energy transfer*

The long-lived carotenoid triplet, T1, forms on the nanosecond timescale and persists beyond the experimental window. Under BChl *a* excitation, where no carotenoid singlet precursor is populated, T1 rises with a single 2100 ps component (AP 800 nm, RP 560 nm). Because this rate is read from the mode specific to this long-lived species, it is free of the spectral overlap that limits TA in the 500 to 560 nm window: the roughly 450 ps component reported earlier [16] reflects an unresolved mixture of overlapping species, whereas the mode-resolved measurement gives 2100 ps directly. This identifies T1 as the product of BChl *a* intersystem crossing followed by triplet-triplet transfer to RG, and is thus the functionally relevant, conventionally photoprotective, carotenoid triplet. Under carotenoid ($S_2$) excitation, the entangled pair decays to the ground state and does not itself generate a separated carotenoid triplet. The long-lived T1 signature nonetheless appears under carotenoid excitation, both in the TA spectra and in the associated FSRRS modes, because about 50% of the $S_2$ population transfers to BChl *a* (*vide supra*); the excited BChl *a* undergoes intersystem crossing to its triplet, which is then transferred to the carotenoid to form T1. Under BChl *a* excitation the same intersystem-crossing and transfer route operates in isolation, and gives the single 2100 ps rise. In both cases T1 is set by BChl *a* intersystem crossing rather than by the carotenoid excited-state manifold. The entangled pair and T1 are therefore mechanistically independent: the pair decays to the ground state, and T1 is populated by triplet transfer from BChl *a*.

### *Heterofission is not supported in LH2*

Wang et al. propose a carotenoid–BChl *a* heterofission pathway in which S2 excitation generates a triplet pair split between the carotenoid and a neighbouring BChl *a* that recombines to $Q_y$ [22]. The evidence does not uniquely support it: the BChl *a* triplet assignment rests on a weak near-infrared feature beneath a $Q_y$ bleach more than 2.5 times stronger, resolved at single wavelengths after global analysis was set aside, and with the near-zero exchange coupling ($J \approx 0$) read as two triplets on different molecules. These observations could be equally reproduced by a weakly-coupled entangled pair $^1$(TT) on a single carotenoid, with triplet contributions at opposite ends of the conjugated backbone with no break in π-conjugation. Our mode-selective probe resolves these assignments. Under carotenoid resonance we recover the fingerprint of a carotenoid-localised entangled pair, and not the carotenoid T1 that a separated pair would leave behind, with no promptly-populated carotenoid triplet proposed by heterofission. The BChl *a* triplet itself lies outside the resonance window of this experiment.

The couplings of this system settle the mechanism. For RG in LH2, the carotenoid – BChl *a* couplings are Coulombic and account quantitatively for the sub-100 fs depopulation of S2, with the S2-$Q_x$-B850 channel dominating and total S2 depopulation near 85 fs [54-55]. Summed over the BChl ring, they correspond to an effective coupling of about 230 cm$^{-1}$ (effective separation 11.7 Å) and drive singlet energy transfer, not triplet-pair formation, leaving no kinetic requirement for a heterofission channel. Heterofission instead requires Dexter-type exchange coupling to a specific BChl *a*, which depends on direct π-orbital overlap and falls off steeply with edge-to-edge separation. The only carotenoid–BChl *a* contact close enough to support exchange is with the B800 monomer, hydrogen-bonded at 3.3 Å; the BChl *a* ring dipoles lie at 14 to 16 Å (Figure 5; [54]), where exchange is negligible. The strong coupling is therefore the wrong type, and the one pigment in orbital contact is the wrong partner for a pathway that recombines to the B850 $Q_y$.

**Table 1.** Coulombic couplings between the RG $S_2 \rightarrow S_0$ transition and neighbouring BChl *a* transitions in *Rbl. acidophilus* LH2.

| **RG $S_2 \rightarrow$ BChl *a*** | **Separation (Å)[a]** | **Förster ($cm^{-1}$)** | **TDC ($cm^{-1}$)[b]** |
|---|---|---|---|
| Nearest ring BChl *a* (B850) $Q_x$ | 14.2 | 130 | 46 |
| Ring BChl *a* (B850) $Q_y$ | 16–17 | 83–87 | 33–104 |
| B800 $Q_y$ | 10.2 | 106 | 45 |
| B800 $Q_x$ | 10.2 | 37 | 19 |

[a] Centre-to-centre distance between the transition-dipole centres.

[b] Transition density cube. Values are magnitudes; signs are given in the original source. Exchange (Dexter) coupling is negligible at all separations [54].

An additional argument against the heterofission description in Wang et al.  is that it treats BChls as localised chromophores, neglecting the excitonic nature of B850 [56]. Singlet fission from a BChl a – carotenoid heterodimer would alter the resonance-enhanced Raman spectrum of the carotenoid partner; while the observed $S^*/^1(TT)$ signals show no perturbation when compared to isolated RG in solvents. In addition, in the case of a strongly-coupled pair, reading $J \approx 0$ as two localised triplets on distinct molecules is in any case not straightforward.

## Conclusions

Using state-selective FSRRS under different resonance conditions, combined with a simultaneous global analysis approach, we have disentangled the vibrational signatures and kinetics of the individual excited-state species contributing to the carotenoid dark-state manifold in LH2. Since all datasets share the same kinetic behaviour, while differing only in the amplitudes of the resonance-enhanced vibrational modes, the analysis resolves overlapping species with a level of discrimination that cannot be achieved from any individual dataset.

Since FSRRS data carry no spectral interference between carotenoid and BChl *a* contributions in the probed window, the growth of the long-lived carotenoid triplet is fitted cleanly, as a single 2100 ps component under BChl *a* excitation. This resolves directly the triplet-triplet transfer channel that TA can only infer, and shows that the conventionally photoprotective carotenoid triplet is produced by triplet-triplet transfer from BChl *a* rather than by singlet fission.

On the other hand, the entangled pair $S^*/^1(TT)$ lives markedly longer in LH2 (about 60 ps) than in solution (about 7 ps), while its vibrational fingerprint, including the newly-resolved $\nu_{1a*}$ mode, is unchanged. We attribute this stabilisation to the backbone twisting imposed by the binding pocket, as proposed elsewhere[8], with a minor contribution due to the high polarisability of the binding pocket. The stabilisation does not open a pathway to separated triplets or to carotenoid-BChl *a* heterofission: the stabilised pair still decays directly to the ground state. Thus, the protein scaffold tunes the lifetime of the entangled pair without changing its fate.

Our results show that the short-lived $S^*/^1(TT)$ entangled pair and the long-lived carotenoid triplet arise from distinct photophysical pathways in purple bacterial LH2: the former undergoes rapid ground-state recovery, whereas the latter results from BChl-to-carotenoid triplet sensitisation and serves as a photoprotective triplet sink for the antenna.

## Experimental Methods

**Sample Preparation.**
Liquid cultures of *Rbl. acidophilus* 10050 were grown anaerobically in the light at 30 °C in Pfenning's medium [57]. Cells were harvested by centrifugation, membranes prepared and the LH2 (B800-850) complexes isolated and purified in the presence of the detergent *N,N*-dimethyldodecylamine *N*-oxide (LDAO) (Fluka) as described previously[58-59]. The purified LH2 antennae were stored in TL buffer (100 mM NaCl, 0.05% (w/v) LDAO, 20 mM Tris.HCl, pH 8.5); the same buffer was used for spectroscopic studies. Rhodopin glucoside was prepared from the same LH2 complexes, as described elsewhere [30].

Steady State Absorption and Raman measurements. Absorption spectra were measured using a Varian Cary E5 scanning spectrophotometer, using a square cell with a 1 cm path length. Resonance Raman spectra were recorded at room temperature with excitations obtained from a Coherent $Ar^+$ (Sabre) laser. Output laser powers of 10–100 mW were attenuated to < 5 mW at the sample. Scattered light was collected at 90° to the incident light, and focused into a Jobin-Yvon U1000 double-grating spectrometer (1800 grooves/mm) equipped with a red-sensitive, back-illuminated, LN2-cooled CCD camera. Sample stability and integrity were assessed based on the stability of the Raman signal.

Time-resolved femtosecond transient absorption (fsTA) and femtosecond stimulated resonance Raman spectroscopy (FSRRS). Samples were transferred into a 1 mm path-length, fused-silica cuvette, adjusted to an OD of 0.6 - 0.8 (for both fsTA and FSRRS) at the actinic pump, and deoxygenated by purging with $N_2$ for 30 min. Sample integrity was monitored by recording the steady-state absorption spectrum before and after each run, and fresh aliquots were used whenever the absorbance changed by more than 10 %. A PHAROS Yb:KGW femtosecond laser (Light Conversion; 1030 nm, 120 fs, 10 kHz, 10 W) was split into three beams to generate the white-light probe (WL), actinic pump (AP) and Raman pump (RP), as described elsewhere [24]. White-light probe (WL): A fraction of the 1030 nm fundamental was attenuated by a neutral-density filter and focused into a 13 mm sapphire plate to generate a broadband continuum (~480–1100 nm). Actinic pump (AP): A separate portion of the fundamental fed an ORPHEUS HE optical parametric amplifier, yielding wavelength-tunable pulses (~150 fs FWHM; ~100 $cm^{-1}$ bandwidth; pump fluence 150 $\mu J\ cm^{-2}$). Raman pump (RP): The remaining beam seeded a second-harmonic band compressor (SHBC), whose output drove an ORPHEUS PS OPA and LYRA difference-frequency stage to deliver a narrowband Raman pump (~5 $cm^{-1}$ resolution; ~3 ps FWHM; pump fluence 400 - 600 $\mu J\ cm^{-2}$). All three beams were directed through independent delay stages into a HARPIA-TA spectrometer. AP and RP beams were chopped at 1 and 0.5 Hz, respectively; the transmitted WL was spatially filtered, collimated, and dispersed by an Andor Kymera 193i spectrograph (Oxford Instruments) onto a 256-pixel Hamamatsu S8380 diode array (200–1100 nm). Broadband TA: 300 g/mm grating (800 nm blaze); $\Delta A(t,\lambda)$ obtained from the integrated intensity difference between 2000 pumped and 2000 unpumped WL pulses per delay. FSRRS: 1200 g/mm grating (600 nm blaze); spectra computed over 5000 WL shots using a four-state chopping scheme: I(RP on, AP on), I(RP off, AP on), I(RP on, AP off), and I(RP off, AP off) [25]. Grating calibration was carried out using carotenoid resonance Raman peaks, and all measurements were performed at room temperature.

Data analysis. Baselines were corrected with a 6/8-order polynomial and datasets were analysed from t = 0 by global sequential analysis [38]. For multi-matrix global fitting, component lifetimes were shared across resonance conditions as described in the main text. The script is available in open access on GitHub [60].

## Associated Content

*Supporting Information.* Transient absorption of LH2 under carotenoid and BChl *a* excitation; TA of RG in THF; FSRRS matrices, global fits, residuals, EADS, and kinetic traces for RG in solution and in LH2 across RP wavelengths and under carotenoid and BChl *a* excitation; averaged EADS with standard deviations.

## Author Information

*Corresponding Authors.*

* Juan Jose Romero. Email : juan-jose.romero@i2bc.paris-saclay.fr

* Bruno Robert. Email : bruno.robert@cea.fr

* Manuel J. Llansola-Portoles. Email : manuel.llansola@cnrs.fr

*Notes.* The authors declare no competing financial interest.

## Acknowledgements

Agence Nationale de la Recherche (ANR) SINGLETFISSION grant ANR-23-CE29-0007 (ML).

Agence Nationale de la Recherche (ANR) FISCIENCY grant ANR-23-CE50-009 (ML).

France 2030 PEPR LUMA program SYNFLUX-LUMICALS grant ANR-23-EXLU-0001(ML).

France 2030 PEPR LUMA program ULTRAFAST platform grant ANR-22-EXLU-0002 (BR).

French Infrastructure for Integrated Structural Biology (FRISBI) grant ANR-10-INSB-05 (I2BC Biophysics Pole) (BR, AG, CI, AP).

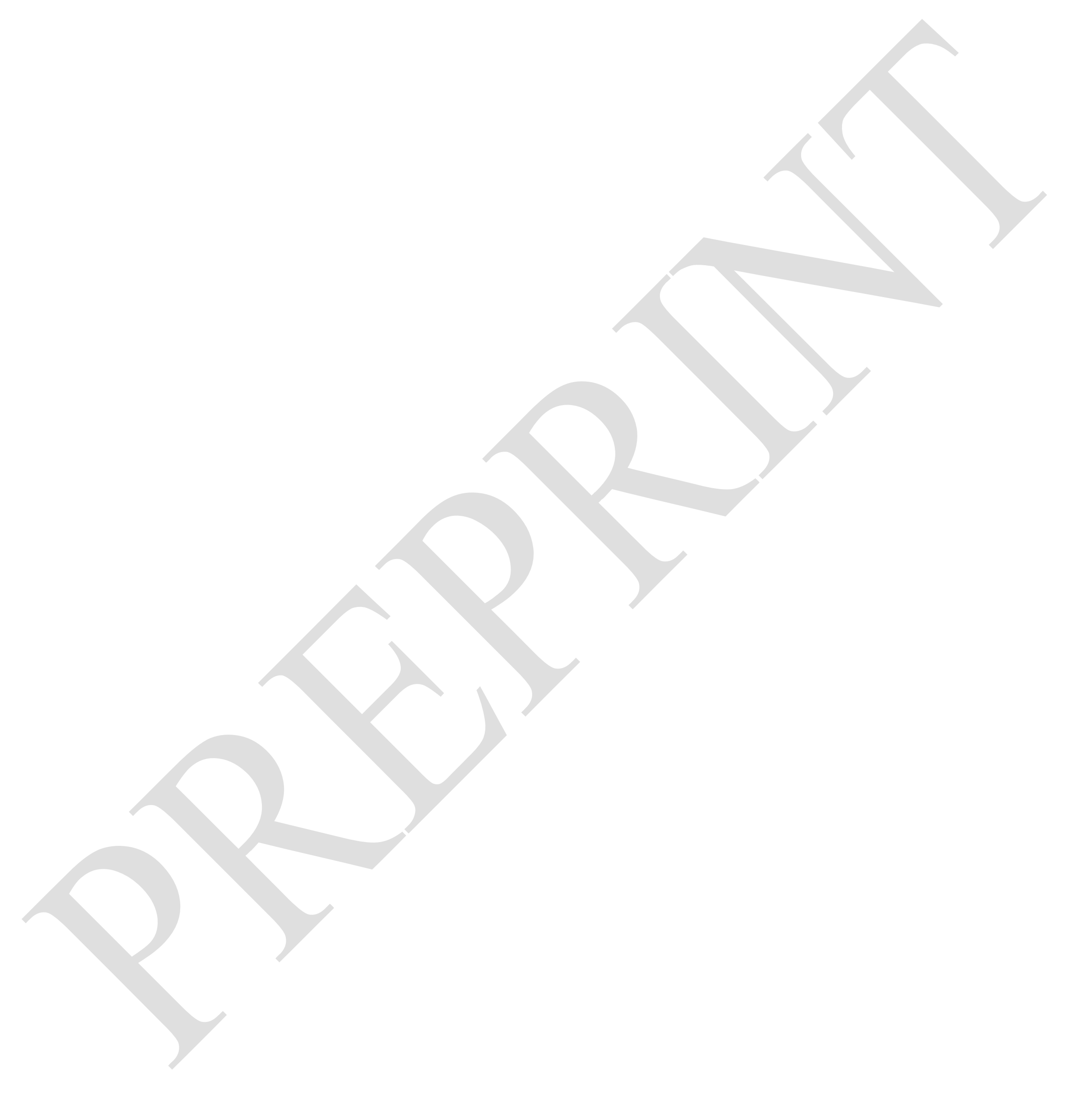

# Supporting Information

## The origin of carotenoid triplets in purple photosynthetic bacteria

Juan J. Romero*, Andrew Gall, Viola D'mello, Cristian Ilioaia, Andrew A. Pascal, Bruno Robert*, Manuel J. Llansola-Portoles*

*Université Paris-Saclay, CEA, CNRS, Institute for Integrative Biology of the Cell (I2BC), 91190 Gif-sur-Yvette, France.*

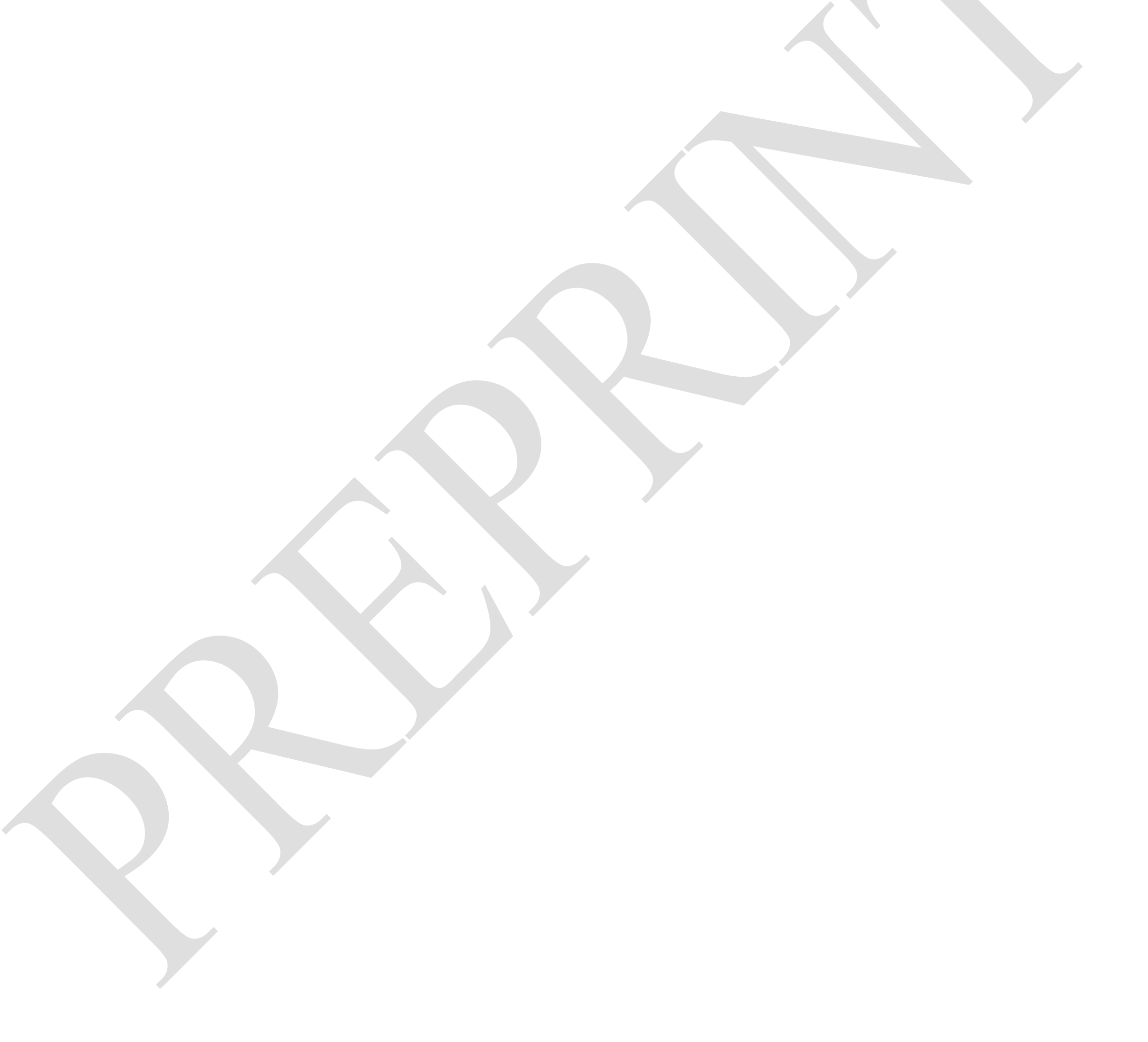

# Multi-Matrix Global Analysis

## Global and target analysis of the resonance-selective FSRRS matrices

The approach for multi-Matrix global Analysis derives from the global- and target-analysis model, and its extension to vibrational spectra shared across several data matrices.

### The bilinear data model

A time-resolved experiment returns a data matrix. The rows index the pump-probe delay $t$ and the columns index the probe coordinate $x$, where $x$ is a wavelength (nm) in transient absorption or a Raman shift ($\mathrm{cm}^{-1}$) in stimulated Raman scattering. We write the signal as $\boldsymbol{\Psi}(\boldsymbol{t},\boldsymbol{x})$, or as an $n_t \times n_x$ matrix $\boldsymbol{\Psi}$.

The analysis rests on one assumption: the signal is bilinear. At every delay the spectrum is a superposition of a small number of components $n_c$, each with a fixed spectrum scaled by a time-dependent population, so the signal at time t and probe coordinate x can be presented as:

$$\Psi(t,x) = \sum_{l=1}^{n_c} c_l\,(t)\,s_l(x), \qquad \text{equivalently} \qquad \boldsymbol{\Psi} = \mathbf{C}\,\boldsymbol{S} \qquad \text{(S1)}$$

Here $\boldsymbol{C} = \boldsymbol{c}^{\boldsymbol{T}}$ is the $n_t \times n_c$ transposition of the concentration matrix, whose column $l$ is the population $c_l(t)$ of the l-th component, and $\boldsymbol{S}$ is the $n_c \times n_x$ spectra matrix, where row l represents the spectrum of the l-th component. This low-rank model is the basis of global and target analysis as formulated for time-resolved spectroscopy by van Stokkum, Larsen and van Grondelle [1]. The probe coordinate $x$ is the only quantity that changes when the technique changes, a point we return to below.

### Separation of the linear and non-linear parameters

The undetermined parameters fall into two classes. The spectra $\boldsymbol{S}$ enter the model linearly. The kinetic quantities, namely the lifetimes or rate constants, the time-zero $t_0$, and the instrument response function (IRF), enter $\boldsymbol{C}$ non-linearly; we collect them in a vector $\boldsymbol{\theta}$, so that $C = C(\theta)$.

For any trial $\boldsymbol{\theta}$**,** the best spectra follow in a single step. Minimising $\|\boldsymbol{\Psi} - \boldsymbol{C}(\boldsymbol{\theta})\,\boldsymbol{S}\|^2$ over $S$ gives the linear least-squares solution

$$\widehat{\boldsymbol{S}}(\boldsymbol{\theta}) = \boldsymbol{C}(\boldsymbol{\theta})^{+}\,\boldsymbol{\Psi} \qquad \text{(S2)}$$

where $C^{+}$ is the Moore-Penrose pseudoinverse. Substituting $\widehat{\boldsymbol{S}}$ back removes the spectra from the problem, so the residual depends on $\theta$ alone,

$$\boldsymbol{R}(\boldsymbol{\theta}) = \boldsymbol{\Psi} - \boldsymbol{C}(\boldsymbol{\theta})\,\boldsymbol{C}(\boldsymbol{\theta})^{+}\,\boldsymbol{\Psi} = [\mathbf{I} - \boldsymbol{C}(\boldsymbol{\theta})\,\boldsymbol{C}(\boldsymbol{\theta})^{+}]\,\boldsymbol{\Psi} \qquad \text{(S3)}$$

The bracketed operator projects the data onto the orthogonal complement of the column space of $\boldsymbol{C}(\boldsymbol{\theta})$. Only the non-linear vector $\boldsymbol{\theta}$ is searched by the optimisation algorithm, and the spectra are reconstructed from $\boldsymbol{\theta}$ at each step. This separation is the variable-projection method of Golub and Pereyra [2]; its application to time-resolved spectra follows van Stokkum and co-workers [1].

The reduction is what makes the fit tractable. A matrix with several hundred probe channels contributes only with its few kinetic parameters to the non-linear search, because every spectrum is recovered analytically from Equation S2. When the populations must be non-negative, the per-channel solution is acquired by non-negative least squares in place of the pseudoinverse. The reduced objective $\|\boldsymbol{R}(\boldsymbol{\theta})\|^2$ is minimised with a trust-region least-squares algorithm.

**The kinetic model and the concentration matrix**

The concentration matrix is not free. Its columns are the solution of a first-order compartmental scheme, namely a set of coupled linear ordinary differential equations,

$$\frac{d\mathbf{c}(t)}{dt} = \boldsymbol{K}\,\mathbf{c}(t), \qquad \mathbf{c}(0\;\;) = \mathbf{j} \qquad \text{(S4)}$$

The vector $\mathbf{c}(t)$ holds the $n_c$ populations, $\mathbf{j}$ is the initial population set by the excitation, and $\boldsymbol{K}$ is the transfer matrix. The off-diagonal elements $K_{ij}$ are the rate constants for transfer from compartment $j$ to compartment $i$. The diagonal elements $K_{jj}$ are the negative sum of all rates leaving compartment $j$, including any decay to a ground state that is not tracked as a compartment.

The scheme is solved by diagonalising $\boldsymbol{K}$. Writing $\boldsymbol{K} = \boldsymbol{V\Lambda V^{-1}}$ with eigenvalues $\lambda_k$ and eigenvectors in the columns of $\boldsymbol{V}$,

$$\mathbf{c}(t) = \sum_k b_k\, e^{\lambda_k t}\,\mathbf{v}_k, \qquad \mathbf{b} = \boldsymbol{V}^{-1}\mathbf{j}. \qquad \text{(S5)}$$

Each eigenvalue is a decay rate with an associated lifetime $\tau_k = -1/\lambda_k$. The finite pulse duration is included by convolving every mode with the IRF. We take the IRF as a Gaussian of full width at half maximum $w$, equivalently of standard deviation $\sigma = w/\left(2\sqrt{2\ln 2}\right)$. The convolution of a decaying exponential with this Gaussian has a closed form,

$$\left(e^{-t/\tau} \otimes IRF\right)(t) = \frac{1}{2}\exp\left(\frac{\sigma^2}{2\tau^2} - \frac{t-t_0}{\tau}\right) erfc\left(\frac{\sigma^2/\tau - (t-t_0)}{\sqrt{2}\,\sigma}\right), \qquad \text{(S6)}$$

so the populations are smooth through time zero and no numerical convolution is required [1]. Lifetimes are optimised as $\ln\tau$, which keeps comparable parameter scales when they span several orders of magnitude.

The compartmental scheme fixes the meaning of the whole fit. The three uses of the framework below differ only in the structure imposed on $\boldsymbol{K}$.

**Sequential scheme and evolution-associated difference spectra (EADS)**

The simplest connected scheme is a unidirectional cascade, $A \to B \to C \to \cdots$, in which all population begins in the first compartment and each compartment decays into the next. The transfer matrix is lower bidiagonal,

$$\boldsymbol{K} = \begin{pmatrix} -k_1 & 0 & 0 & \\ k_1 & -k_2 & 0 & \\ 0 & k_2 & -k_3 & \\ & & \ddots & \ddots \end{pmatrix}, \qquad k_l = 1/\tau_l, \qquad \text{(S7)}$$

with $\mathbf{j} = (1,0,\ldots,0)^{\mathsf{T}}$. Solving Equation S1 with these populations yields one spectrum per compartment - these are the evolution-associated difference spectra (EADS). Each EADS is the observed spectrum when its compartment is maximally populated, and the ordered set shows how the global spectrum evolves as population flows down the cascade. The fitted $\tau_l$ are the successive time constants of that evolution.

The sequential model commits to nothing beyond the ordering of the lifetimes. It is the standard first description of a dataset: it summarises the kinetics with the fewest components and returns lifetimes and spectral shapes that any mechanistic model must then reproduce.

**Target scheme and species-associated spectra (SAS)**

Target analysis replaces the cascade with a specific kinetic scheme. The scheme states which compartments are connected, in which direction, and with what initial populations; each connection is a free rate constant, and decay to the ground state enters as a loss term on the diagonal of $\boldsymbol{K}$. This defines a general transfer matrix, not restricted to the bidiagonal form of Equation S7.

Solving Equation S1 with this $\boldsymbol{K}$ gives the species-associated spectra (SAS). Each spectrum now belongs to a physical species of the assumed scheme, and each population $c_l(t)$ is the concentration of that species. The SAS are defined relative to the scheme: changing the connections within the scheme will give different SAS from the same data. The fit therefore tests a mechanism rather than describing the data. A scheme is supported when it reproduces $\boldsymbol{\Psi}$ with SAS that are physically admissible, for instance non-negative where the spectroscopy requires it.

**Relationship between the spectral representations**

The spectral representations are linear combinations of one another. Because the experimentally observed temporal components are the convolution of the kinetic eigenvalues with the IRF$e^{\lambda_k t} \otimes IRF$, they form a common time basis. Then, the data can be written in the species basis or in the eigenvector basis. The two are connected by

$$\boldsymbol{\Psi} = \boldsymbol{C}\boldsymbol{S} = M\,S \qquad M = diag\left(\mathbf{b}e^{\lambda_k t}\right)\boldsymbol{V}^{\mathbf{T}} \qquad \text{(S8)}$$

with $\mathbf{b}$ and $\boldsymbol{V}$ from Equation S5. The decay-associated spectra (DAS) represent a parallel, kinetically independent description in the eigenvector basis, and the SAS or EADS of a compartmental description are thus related by an invertible matrix built from the same eigenvectors and initial populations. A single fit can be reported in whichever representation is appropriate: DAS to display the raw decay components, EADS to show the spectral evolution, or SAS to read the spectra of the postulated species.

**Coherent artefact**

Near time zero, the pump and probe overlap in the sample and produce a non-population signal, the coherent artefact, from cross-phase modulation and related effects. It carries no lifetime and is not a kinetic component. We describe it by appending one or two fixed temporal shapes to the concentration matrix: a Gaussian centred at $t_0$ with the width of the IRF,

$$a_0(t) = \exp\left(-\frac{1}{2}\left[\frac{t-t_0}{\sigma_a}\right]^2\right), \qquad a_1(t) = -\frac{t-t_0}{\sigma_a}\,a_0(t), \qquad \text{(S9)}$$

the second being its first time derivative, which captures the dispersive, sign-changing part of the artefact. These columns are added to $C$ and their spectra are recovered by the same linear solution as every other component (Equation S2), so the artefact is separated from the kinetics without any additional non-linear amplitude parameter.

**Extension to vibrational spectra: modes shared across matrices**

The framework so far is indifferent to the probe coordinate. Replacing wavelength by Raman shift carries global and target analysis unchanged from electronic transient absorption to femtosecond stimulated Raman spectroscopy (FSRS) and its resonance variant (FSRRS): $\boldsymbol{\Psi}(t,\tilde{\nu})$ is still bilinear, the kinetics still reside in $\boldsymbol{C}$, and the vibrational spectra reside in $\boldsymbol{S}$.

The extension we introduce uses a property that is specific to the vibrational observable. The Raman spectrum of a species is a fingerprint of its nuclear structure: the positions and relative intensities of its bands are set by its normal modes, and they do not change between measurements as long as the species is the same. When several matrices are recorded on the same system, for example at different excitation conditions or concentrations, the vibrational modes of each species are therefore common to all of them, and only an overall amplitude differs between matrices.

We impose this directly. Let $\boldsymbol{\Psi_i}$, $i = 1, \dots, M$, be the set of matrices, sharing one kinetic model $\boldsymbol{\theta}$. For a species $c$ whose spectrum is linked across matrices, its spectrum in matrix $i$ is

$$\boldsymbol{S}_{i,c}(\tilde{\nu}) = \alpha_{i,c}\,\boldsymbol{S}_c(\tilde{\nu}), \qquad \alpha_{1,c} = 1 \qquad \text{(S10)}$$

where $\boldsymbol{S_c}$ is a single shared band shape and $\alpha_{i,c}$ is a scalar amplitude for matrix $i$. The gauge $\alpha_{1,c} = 1$ fixes the overall scale, so $\boldsymbol{S_c}$ is defined uniquely. The shared shape is a rank-one factor: one spectral profile common to every matrix, multiplied by a per-matrix number. The objective summed over the matrices,

$$\chi^2\big(\boldsymbol{\theta}, \{\boldsymbol{S_c}\}, \{\alpha_{i,c}\}\big) = \sum_{i=1}^{M} \lVert \boldsymbol{\Psi}_i - \boldsymbol{C}_i(\boldsymbol{\theta})\,\boldsymbol{S_i} \rVert^2\,, \qquad \text{(S11)}$$

is minimised by the same separation as before. For each trial $\boldsymbol{\theta}$, the shared shapes and the amplitudes are the linear unknowns, recovered together in one least-squares solution, and only $\boldsymbol{\theta}$ is searched non-linearly.

Sharing the shapes has two effects. It ties the matrices together through one set of spectra and lifetimes, so that a weak or short-lived species that is poorly constrained in one matrix borrows definition from the others. It also removes the freedom for a species to adopt a different shape in each matrix, which is the freedom that allows noise to imitate spectral change. The link can be restricted to a chosen range of $\tilde{\nu}$, leaving windows where a feature is genuinely matrix-dependent, such as a ground-state bleach, free to differ per matrix; outside those windows the shared fingerprint holds.

The assumption is that the vibrational fingerprint is invariant across the linked matrices. This holds when the same electronic resonance is addressed and the species are unchanged. Where the resonance condition itself is varied, relative band intensities can shift, and the link is then applied only over the modes for which invariance is justified.

## Transient absorption in the fs-to-ns time range

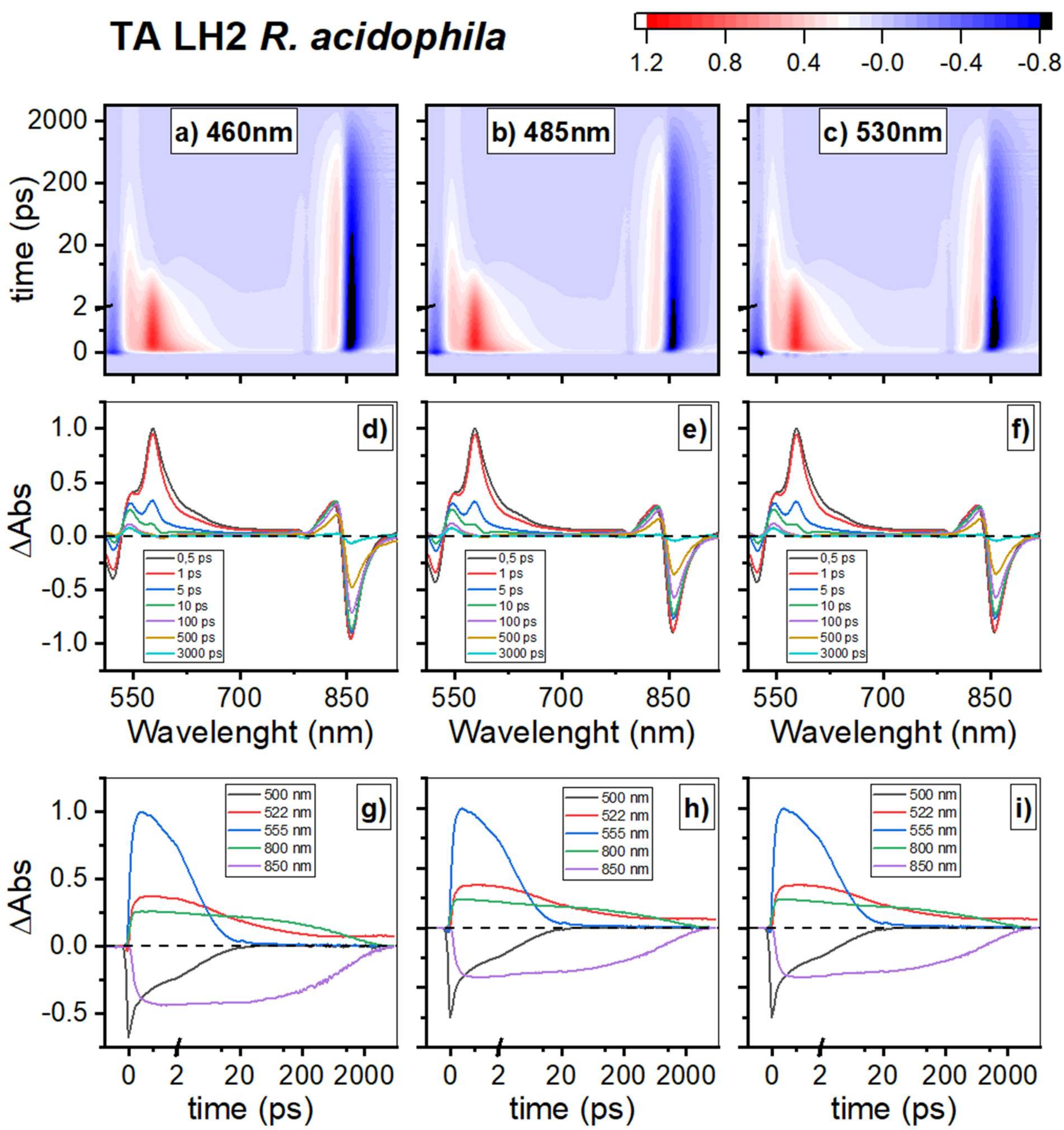

**Figure S1 |** Transient absorption of LH2 upon carotenoid excitation at 460, 485, and 530 nm. (a-c) Raw ΔOD(λ,t) maps for 460, 485, and 530 nm, respectively. (d-f) Time-gated spectra for the same three excitation wavelengths. (g-i) Selected kinetic traces.

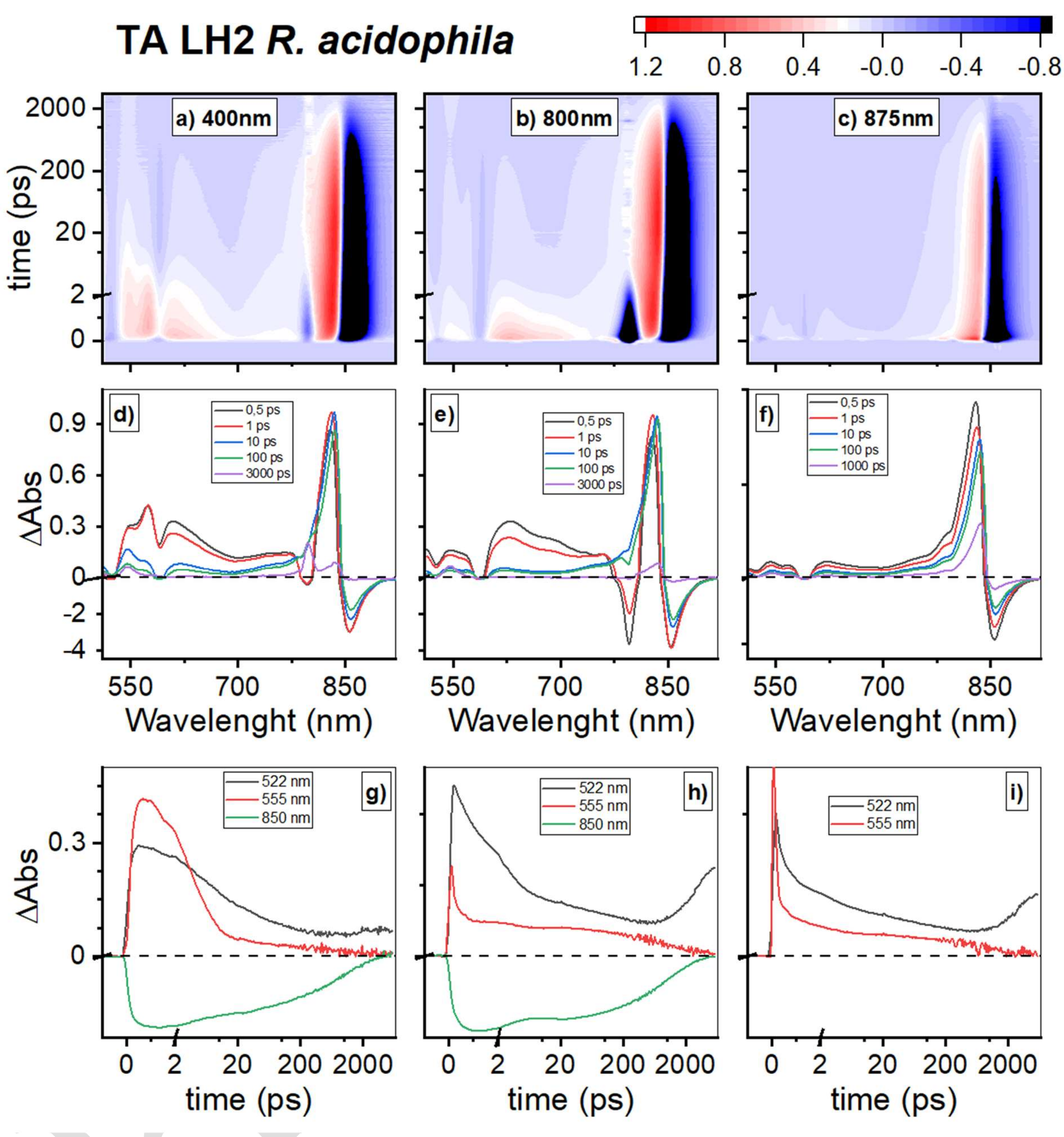

**Figure S2 |** Transient absorption of LH2 upon BChl *a* excitation at 400 nm (Soret), 800 nm, and 875 nm. (a-c) Raw ΔOD(λ,t) maps for 400, 800, and 875 nm, respectively. (d-f) Time-gated spectra for the same three excitation wavelengths. (g-i) Selected kinetic traces.

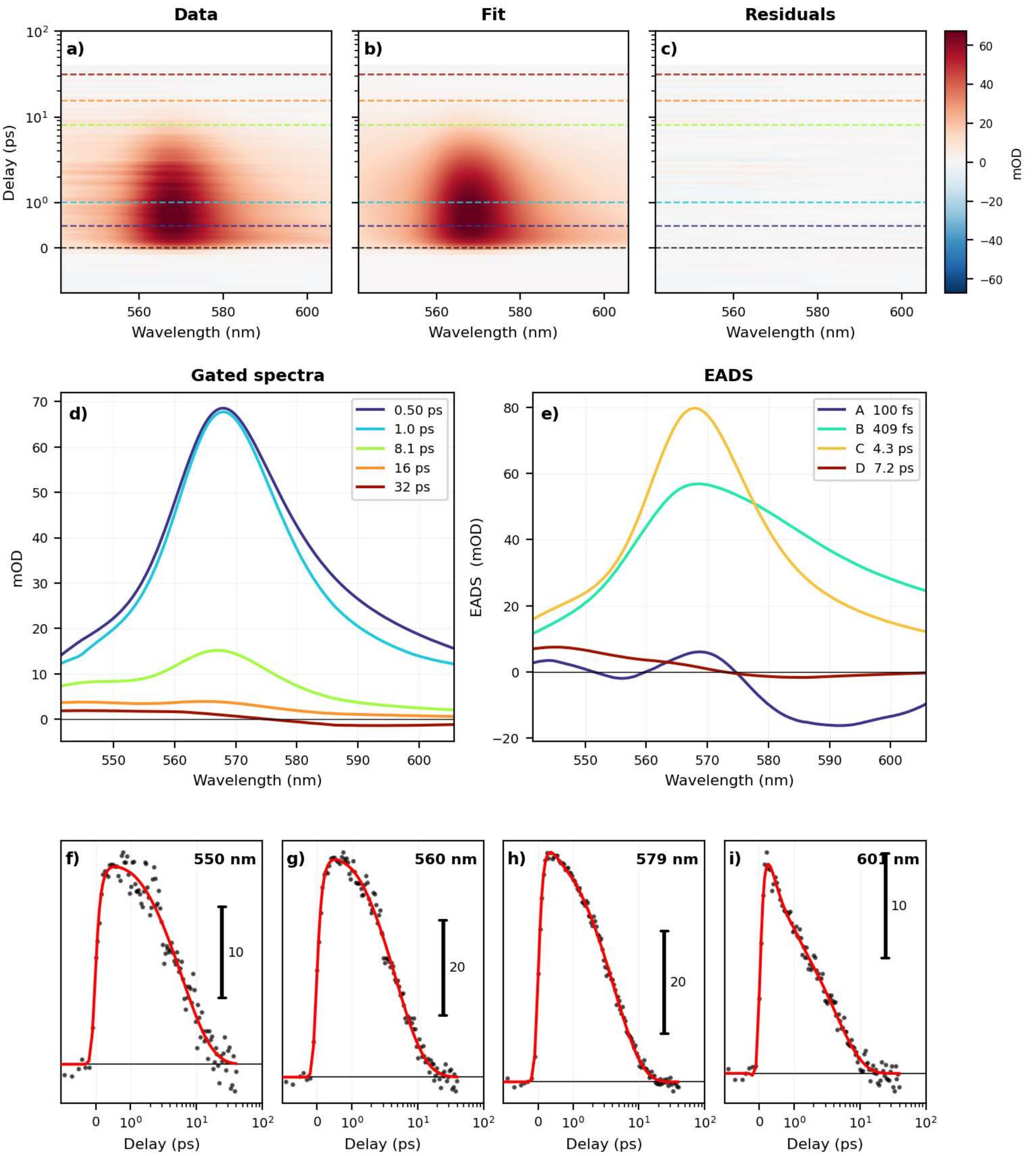

**Figure S3 |** Transient absorption of RG in THF, 485 nm actinic pump. (a) Raw ΔOD(λ,t) map. (b) Global sequential fit (four components). (c) Fit residuals. (d) Time-gated spectra. (e) Evolution-associated difference spectra (EADS) with associated lifetimes. (f-i) Kinetic traces (black dots) with sequential-model fits (red line).

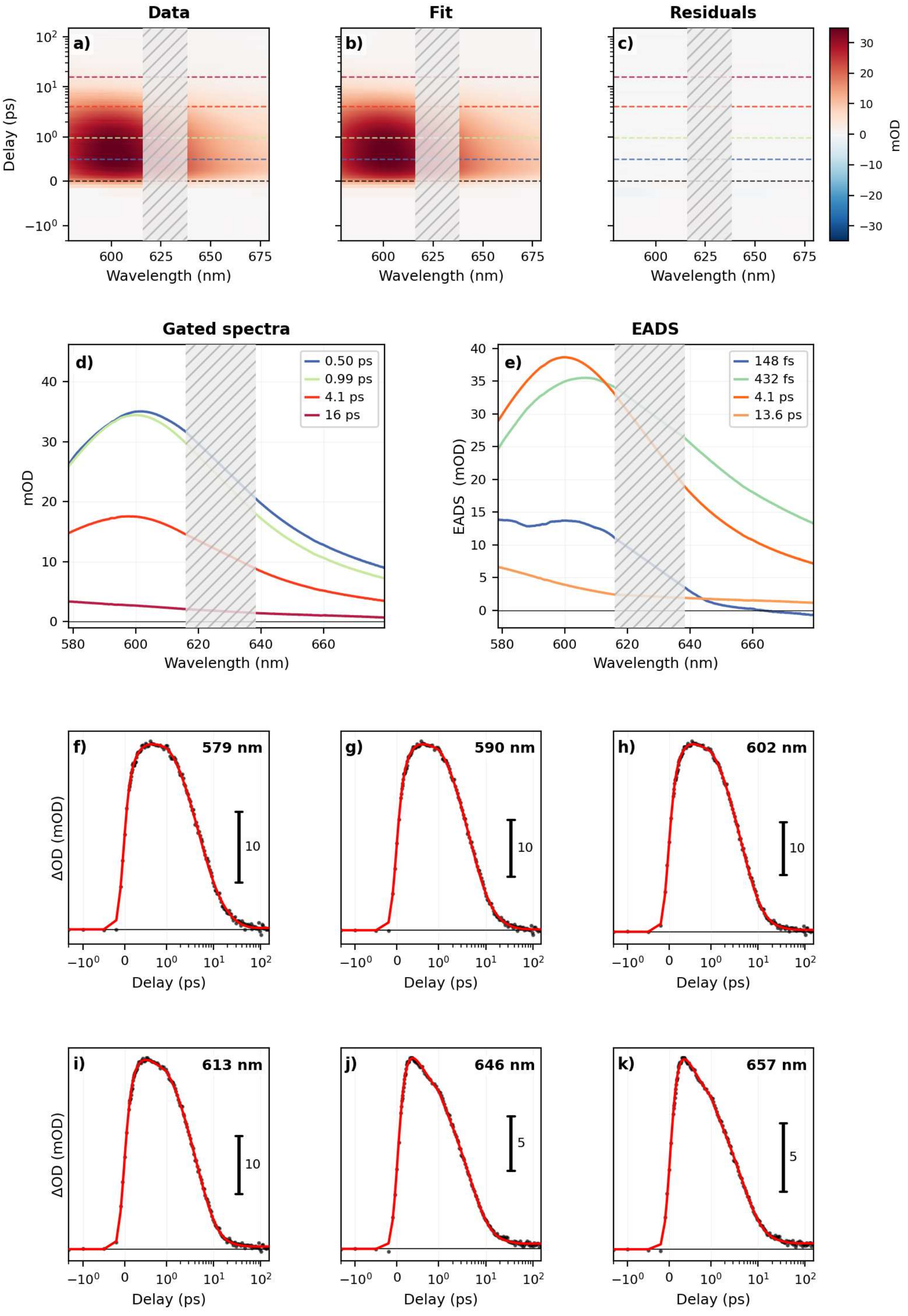

**Figure S4 |** Transient absorption of RG in CS2, 485 nm actinic pump. (a) Raw ΔOD(λ,t) map. (b) Global sequential fit (four components). (c) Fit residuals. (d) Time-gated spectra. (e) Evolution-associated difference spectra (EADS) with associated lifetimes. (f-i) Kinetic traces (black dots) with sequential-model fits (red line). Gray shaded region was slightly distorted by white-light probe and was excluded from global analysis.

## Resonance Raman Spectroscopy

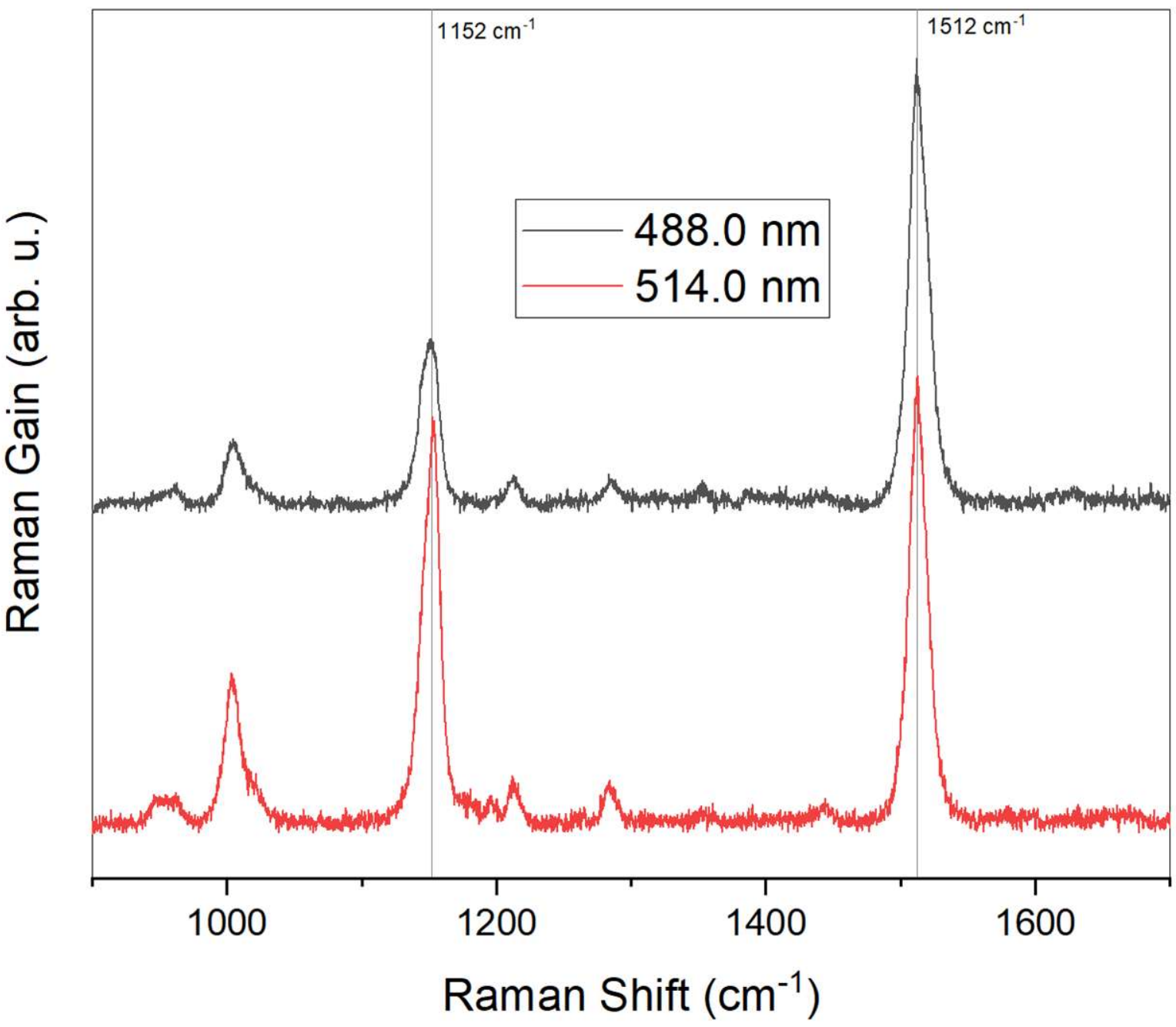

**Figure S5 |** Resonance Raman of LH2 upon excitation at 488.0 and 514.5 nm at room temperature.

## Femtosecond Stimulated Resonance Raman Spectroscopy

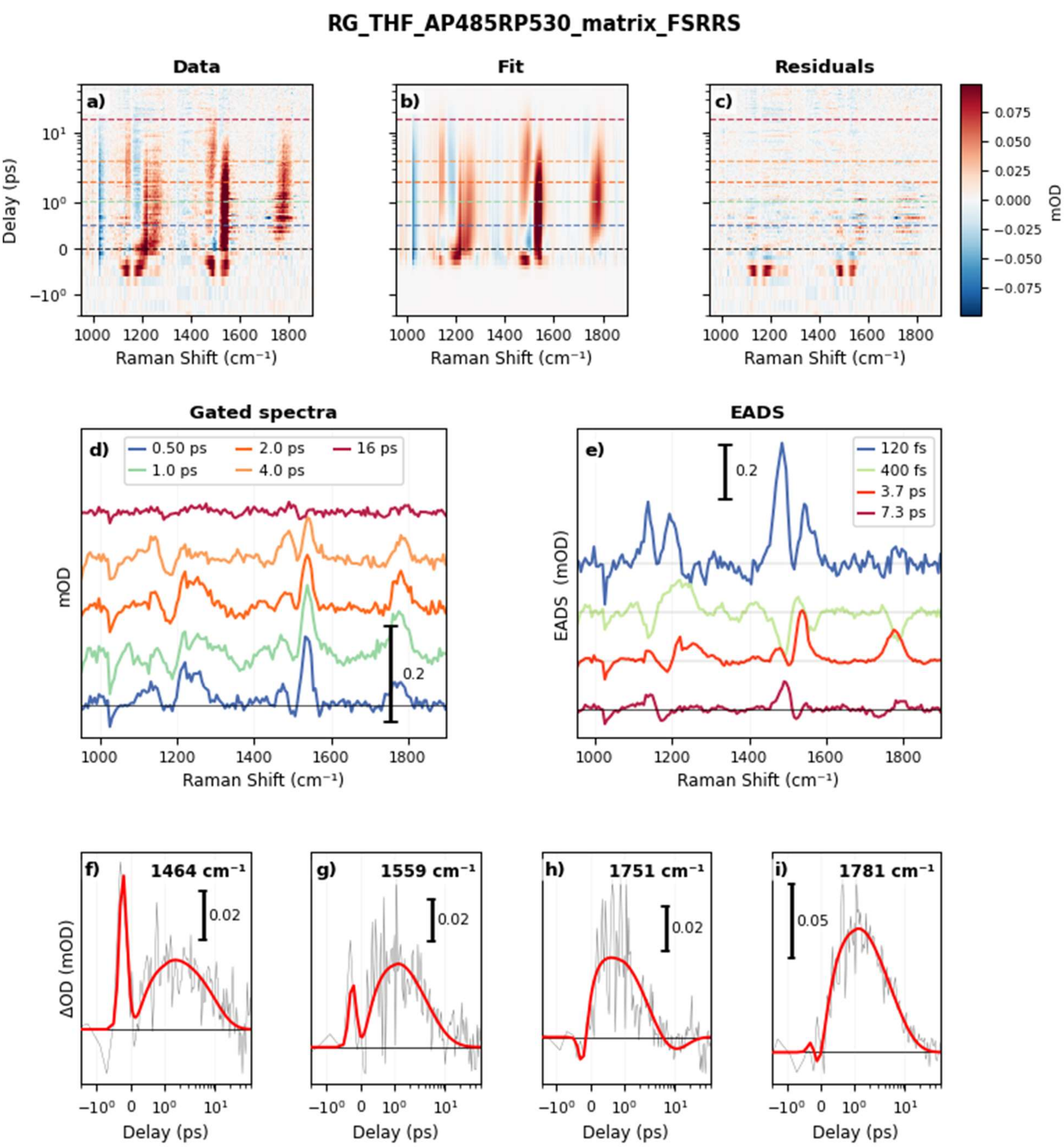

**Figure S6 |** FSRRS of RG in THF, actinic pump 485 nm, Raman pump 530 nm. (a) Raw Raman-gain map as a function of Raman shift ($cm^{-1}$) and pump-probe delay (ps). (b) Global sequential fit of the dataset in (a) with a four-component kinetic model. (c) Fit residuals. (d) Time-gated Raman spectra at selected pump-probe delays. (e) Evolution-associated difference spectra (EADS), along with associated component lifetimes. (f-i) Kinetic traces (gray line) of selected vibrational modes, with the sequential-model fit overlaid (red line).

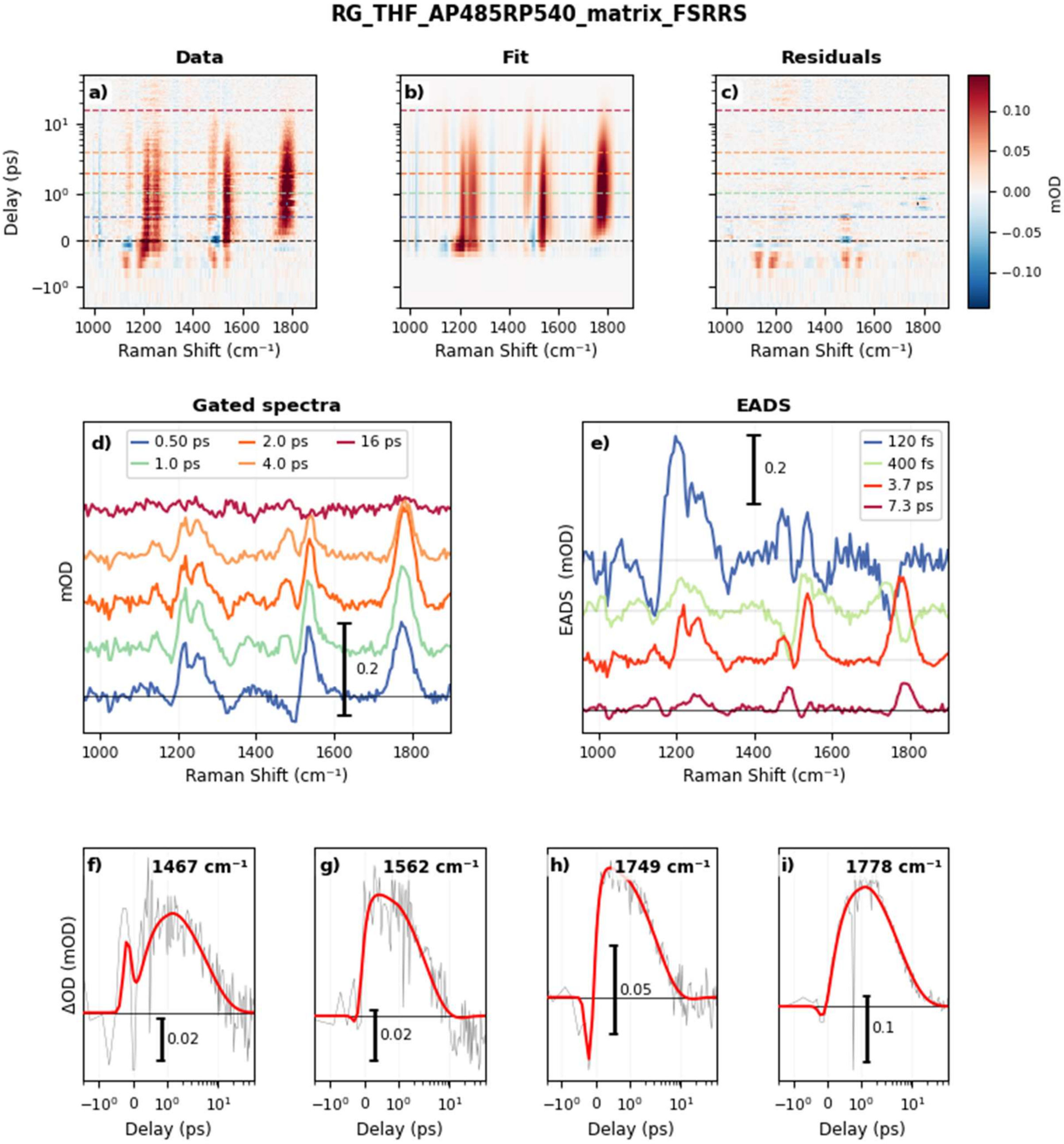

**Figure S7 |** FSRRS of RG in THF, actinic pump 485 nm, Raman pump 540 nm. (a) Raw Raman-gain map as a function of Raman shift ($cm^{-1}$) and pump-probe delay (ps). (b) Global sequential fit of the dataset in (a) with a four-component kinetic model. (c) Fit residuals. (d) Time-gated Raman spectra at selected pump-probe delays. (e) Evolution-associated difference spectra (EADS) with the associated component lifetimes. (f-i) Kinetic traces (gray line) of selected vibrational modes, with the sequential-model fit overlaid (red line).

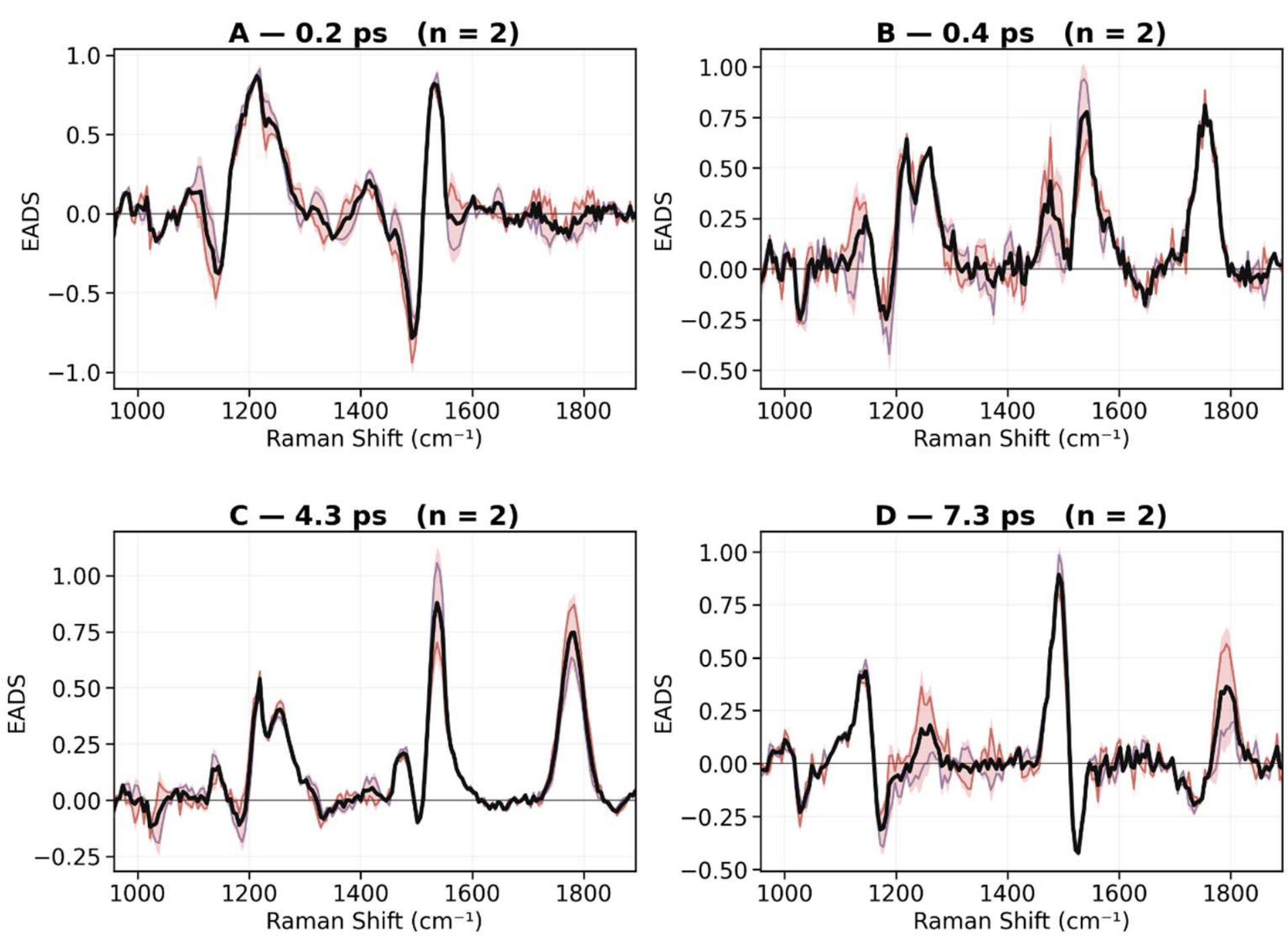

**Figure S8 |** Averaged EADS from multi-matrix global fitting of RG in THF (Raman pump 530 and 540 nm). Mean EADS obtained across the two datasets; the shaded red band shows the standard deviation on the S* ($\nu_{1d}$) component.

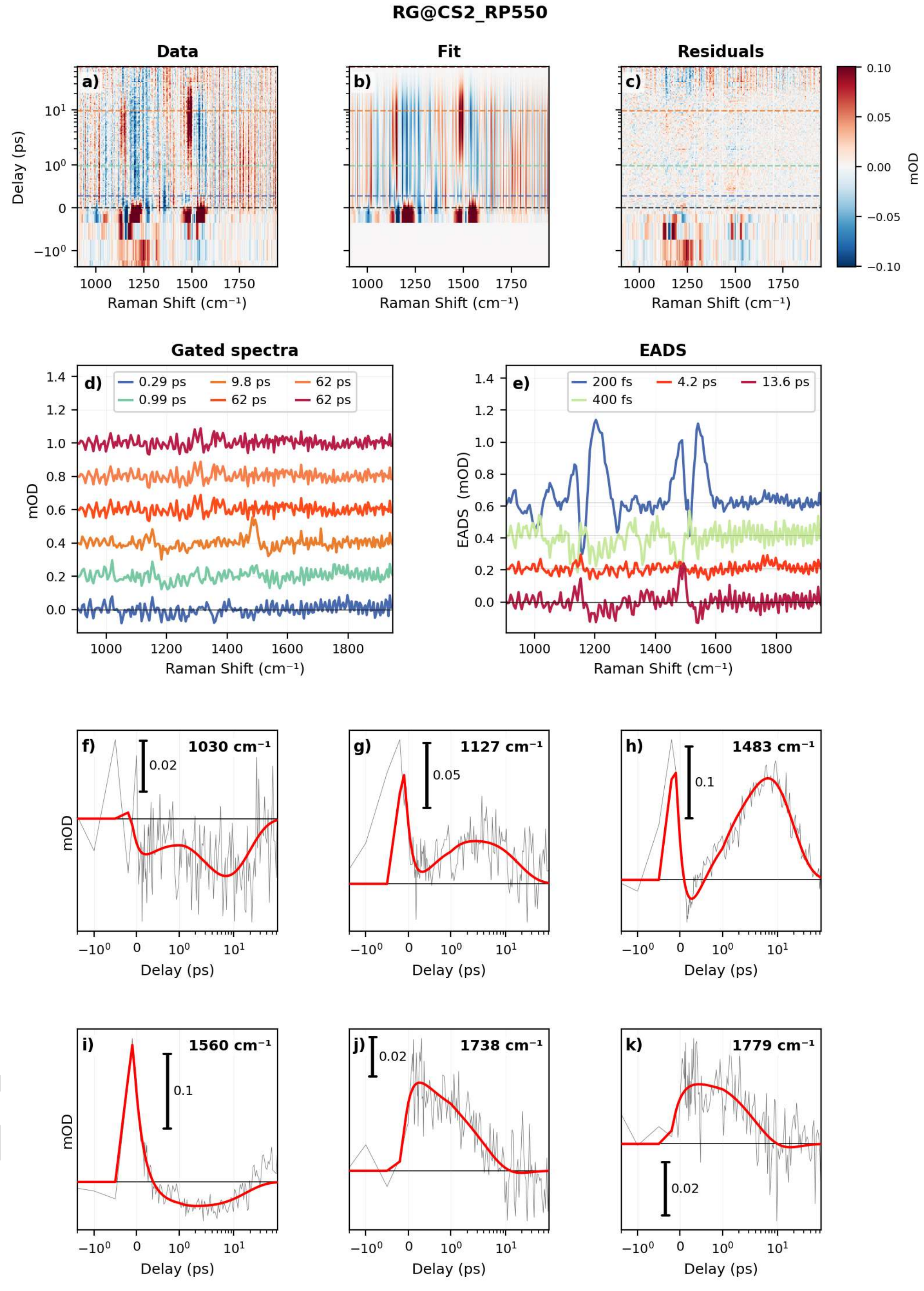

**Figure S9 |** FSRRS of RG in CS2, actinic pump 485 nm, Raman pump 550 nm. (a) Raw Raman-gain map as a function of Raman shift ($cm^{-1}$) and pump-probe delay (ps). (b) Global sequential fit of the dataset in (a) with a four-component kinetic model. (c) Fit residuals. (d) Time-gated Raman spectra at selected pump-probe delays. (e) Evolution-associated difference spectra (EADS) with the associated component lifetimes. (f-i) Kinetic traces (gray line) of selected vibrational modes, with the sequential-model fit overlaid (red line).

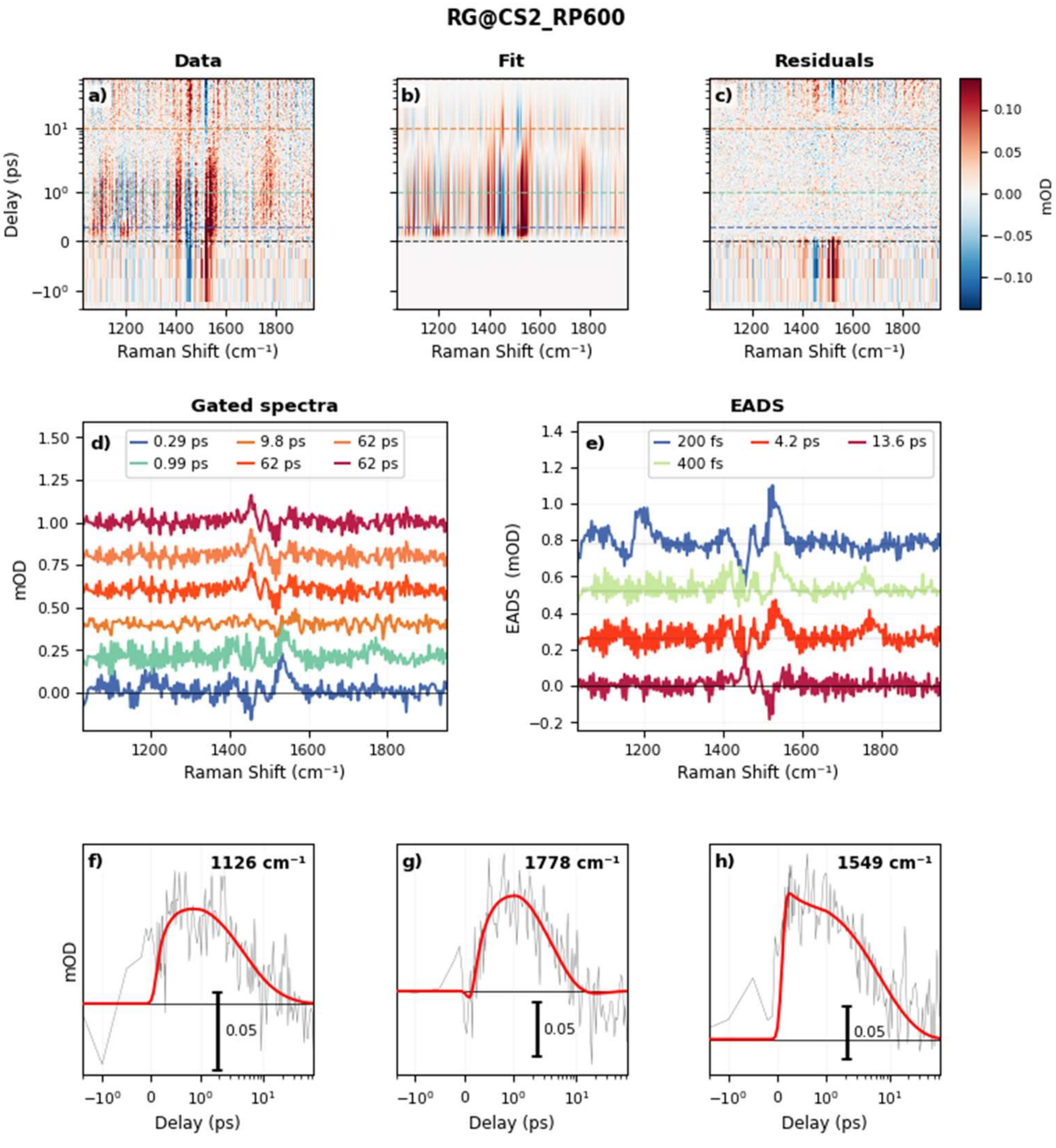

**Figure S10 |** FSRRS of RG in CS2, actinic pump 485 nm, Raman pump 600 nm. (a) Raw Raman-gain map as a function of Raman shift ($cm^{-1}$) and pump-probe delay (ps). (b) Global sequential fit of the dataset in (a) with a four-component kinetic model. (c) Fit residuals. (d) Time-gated Raman spectra at selected pump-probe delays. (e) Evolution-associated difference spectra (EADS) with the associated component lifetimes. (f-i) Kinetic traces (gray line) of selected vibrational modes, with the sequential-model fit overlaid (red line).

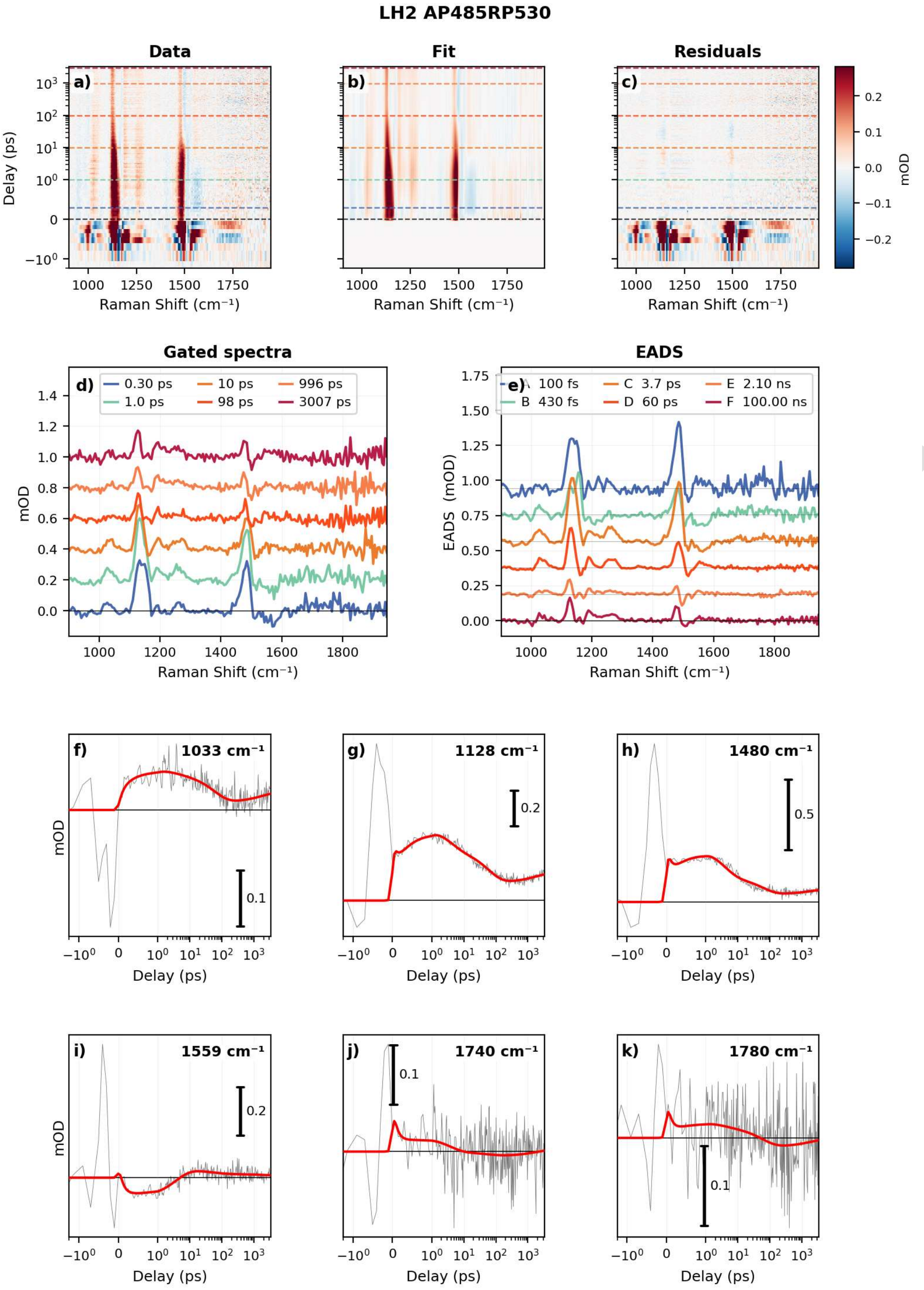

**Figure S11 |** FSRRS of LH2, actinic pump 485 nm, Raman pump 530 nm. (a) Raw Raman-gain map as a function of Raman shift ($cm^{-1}$) and pump-probe delay (ps). (b) Global sequential fit of the dataset in (a) with a six-component kinetic model. (c) Fit residuals. (d) Time-gated Raman spectra at selected pump-probe delays. (e) Evolution-associated difference spectra (EADS) with the associated component lifetimes. (f-i) Kinetic traces (gray line) of selected vibrational modes, with the sequential-model fit overlaid (red line).

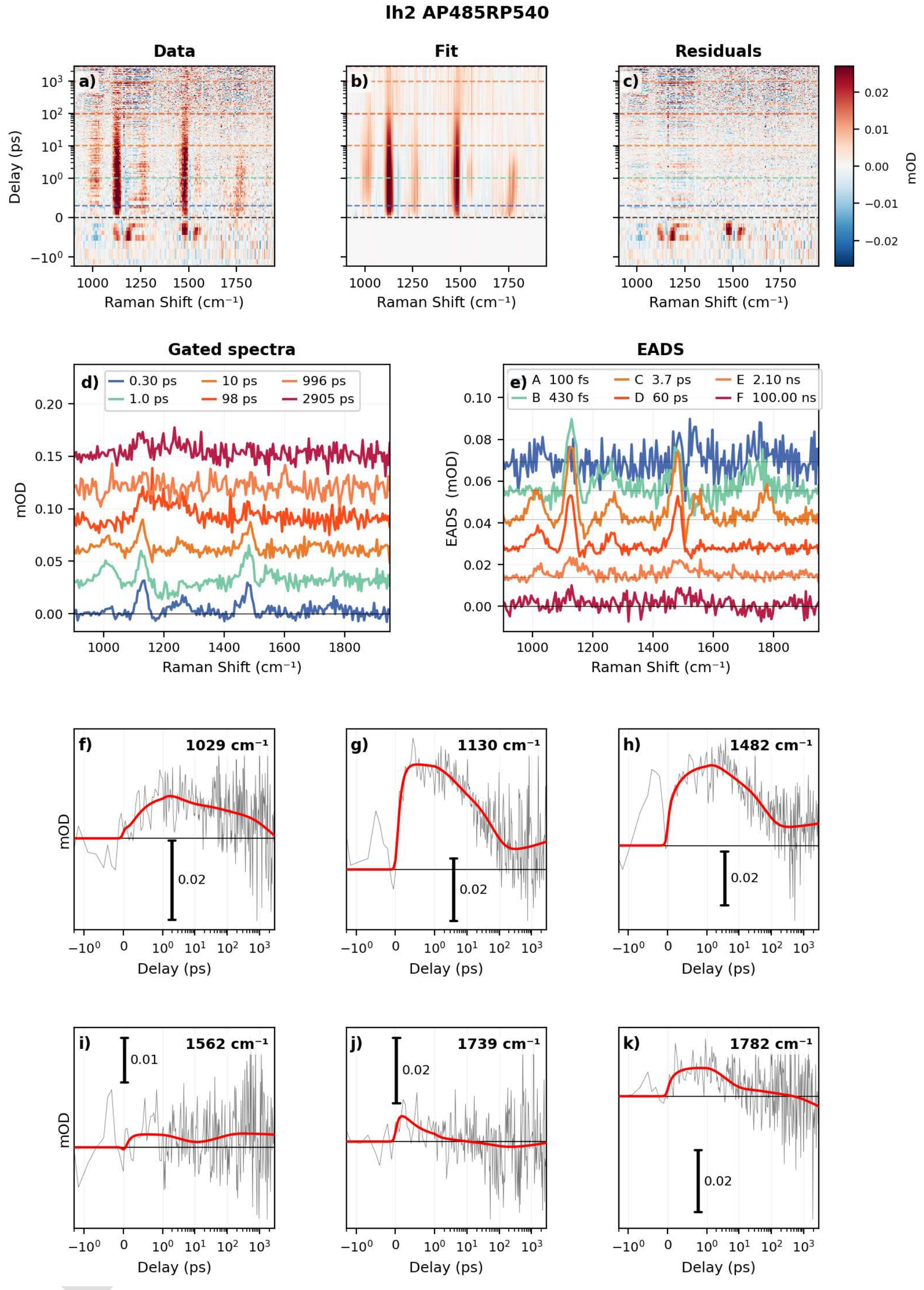

**Figure S12 |** FSRRS of LH2, actinic pump 485 nm, Raman pump 540 nm. (a) Raw Raman-gain map as a function of Raman shift ($cm^{-1}$) and pump-probe delay (ps). (b) Global sequential fit of the dataset in (a) with a six-component kinetic model. (c) Fit residuals. (d) Time-gated Raman spectra at selected pump-probe delays. (e) Evolution-associated difference spectra (EADS) with the associated component lifetimes. (f-i) Kinetic traces (gray line) of selected vibrational modes, with the sequential-model fit overlaid (red line).

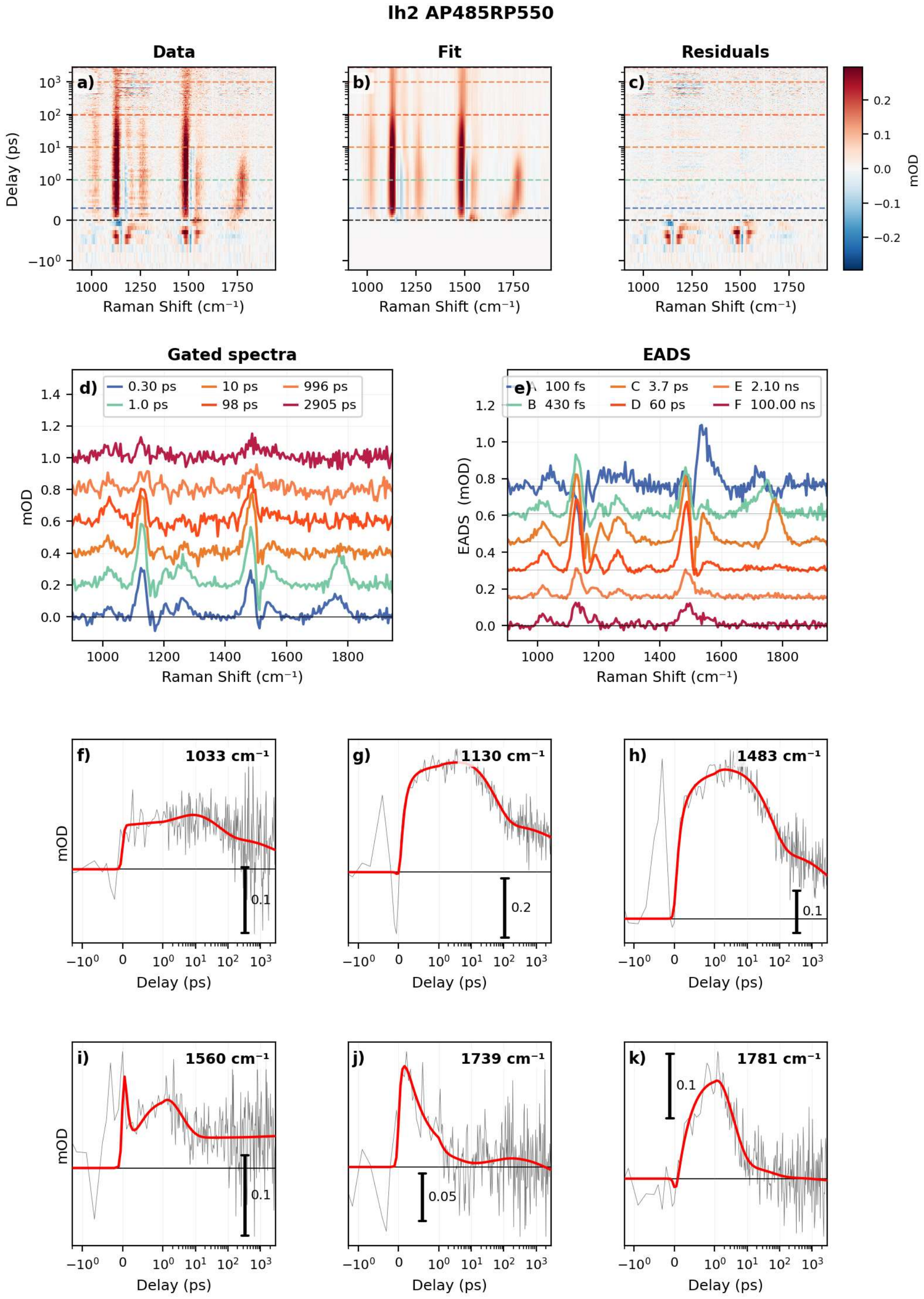

**Figure S13 |** FSRRS of LH2, carotenoid excitation, actinic pump 485 nm, Raman pump 550 nm. (a) Raw Raman-gain map as a function of Raman shift ($cm^{-1}$) and pump-probe delay (ps). (b) Global sequential fit of the dataset in (a) with a six-component kinetic model. (c) Fit residuals. (d) Time-gated Raman spectra at selected pump-probe delays. (e) Evolution-associated difference spectra (EADS) with the associated component lifetimes. (f-i) Kinetic traces (gray line) of selected vibrational modes, with the sequential-model fit overlaid (red line).

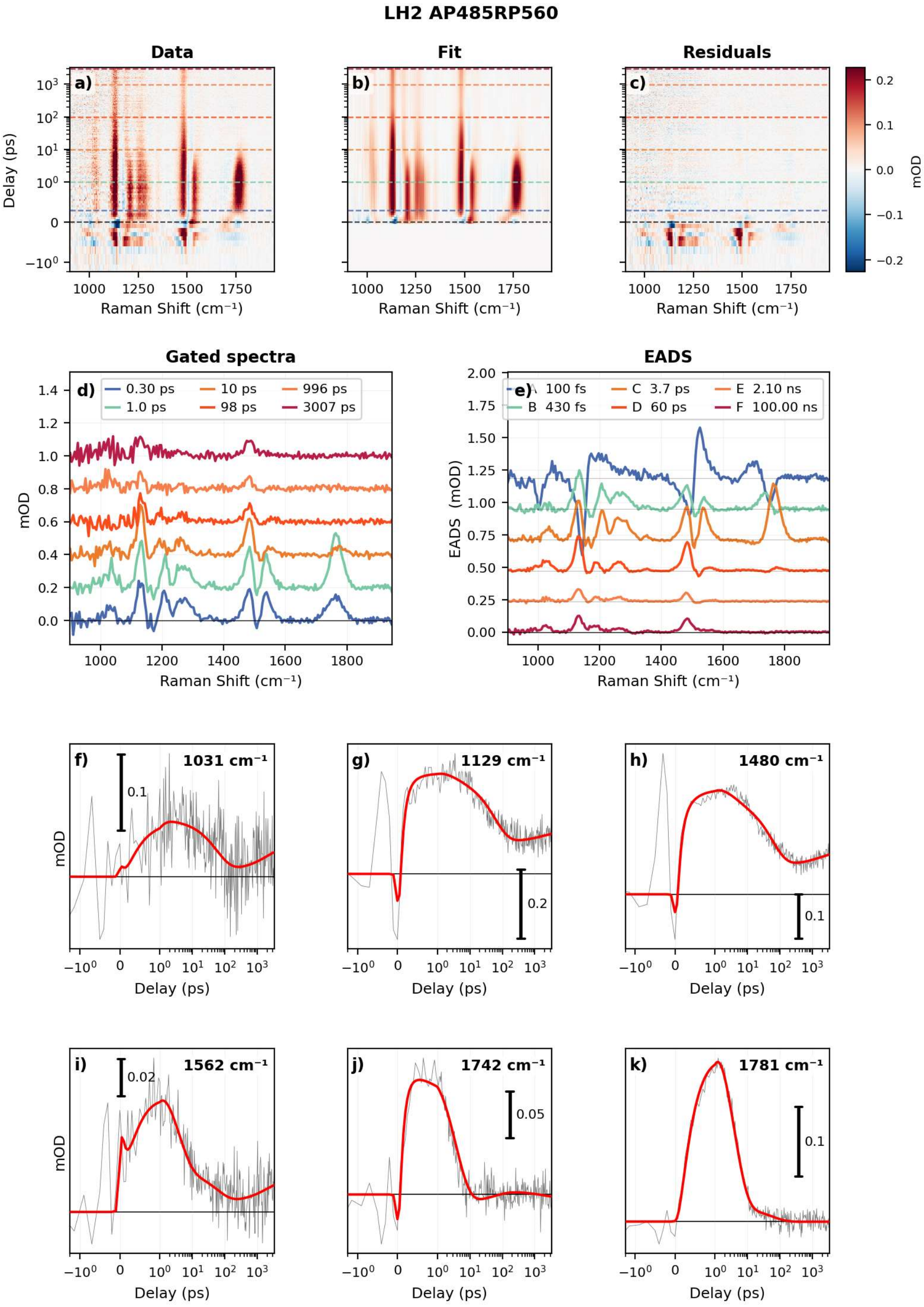

**Figure S14 |** FSRRS of LH2, carotenoid excitation, actinic pump 485 nm, Raman pump 560 nm. (a) Raw Raman-gain map as a function of Raman shift (cm$^{-1}$) and pump-probe delay (ps). (b) Global sequential fit of the dataset in (a) with a six-component kinetic model. (c) Fit residuals. (d) Time-gated Raman spectra at selected pump-probe delays. (e) Evolution-associated difference spectra (EADS) with the associated component lifetimes. (f-i) Kinetic traces (gray line) of selected vibrational modes, with the sequential-model fit overlaid (red line).

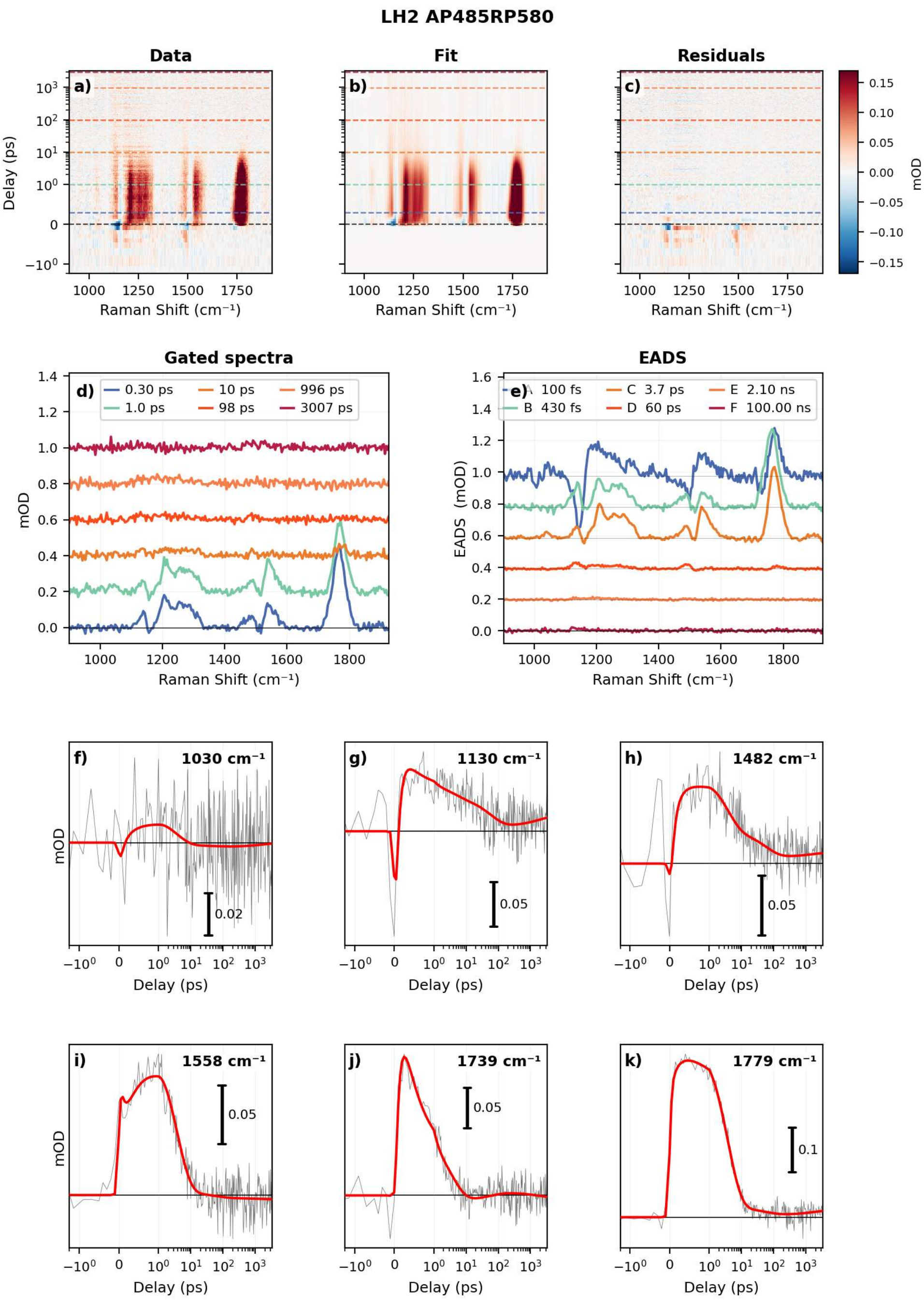

**Figure S15 |** FSRRS of LH2, carotenoid excitation, actinic pump 485 nm, Raman pump 580 nm. (a) Raw Raman-gain map as a function of Raman shift ($cm^{-1}$) and pump-probe delay (ps). (b) Global sequential fit of the dataset in (a) with a six-component kinetic model. (c) Fit residuals. (d) Time-gated Raman spectra at selected pump-probe delays. (e) Evolution-associated difference spectra (EADS) with the associated component lifetimes. (f-i) Kinetic traces (gray line) of selected vibrational modes, with the sequential-model fit overlaid (red line).

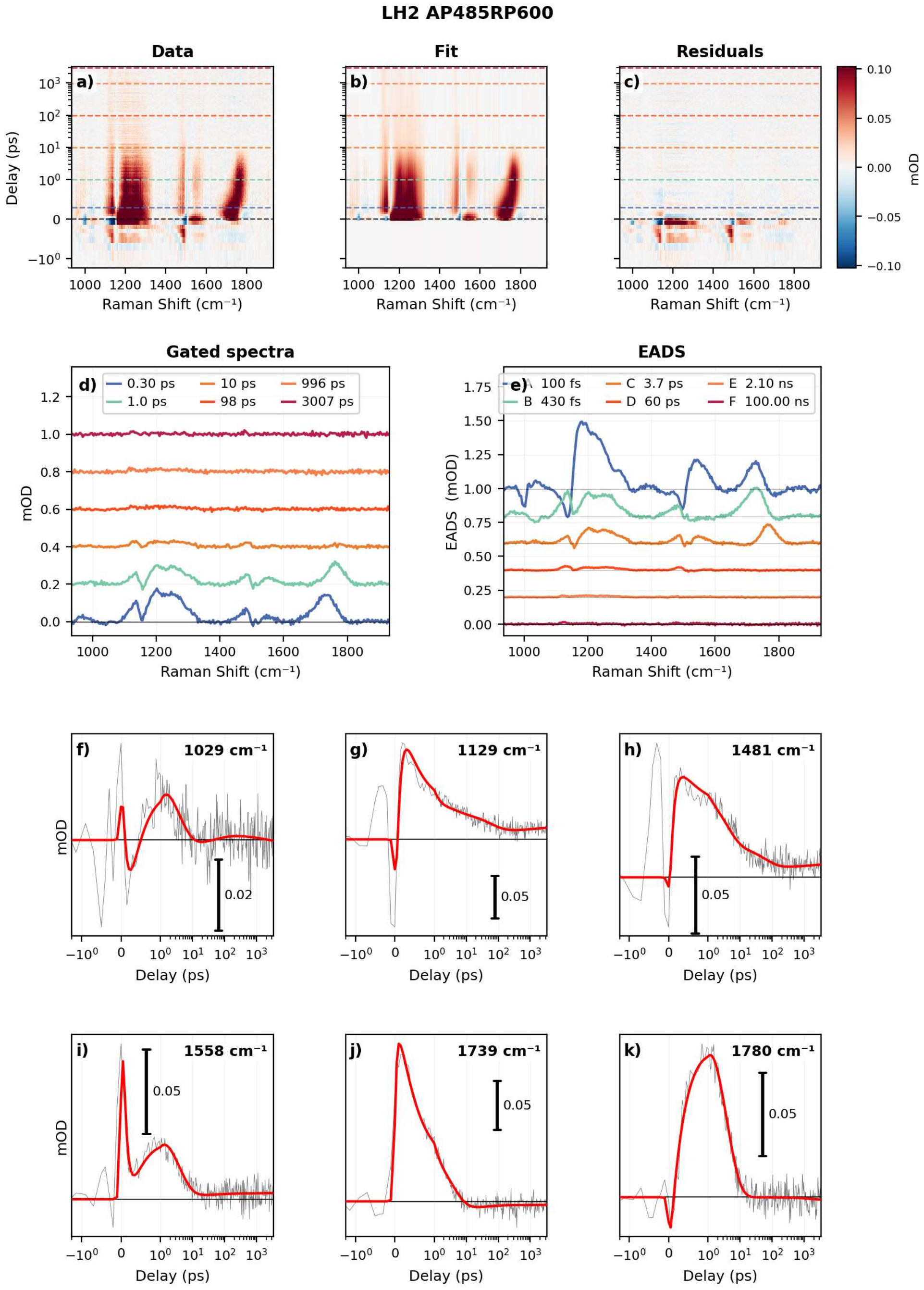

**Figure S16 |** FSRRS of LH2, carotenoid excitation, actinic pump 485 nm, Raman pump 600 nm. (a) Raw Raman-gain map as a function of Raman shift ($cm^{-1}$) and pump-probe delay (ps). (b) Global sequential fit of the dataset in (a) with a six-component kinetic model. (c) Fit residuals. (d) Time-gated Raman spectra at selected pump-probe delays. (e) Evolution-associated difference spectra (EADS) with the associated component lifetimes. (f-i) Kinetic traces (gray line) of selected vibrational modes, with the sequential-model fit overlaid (red line).

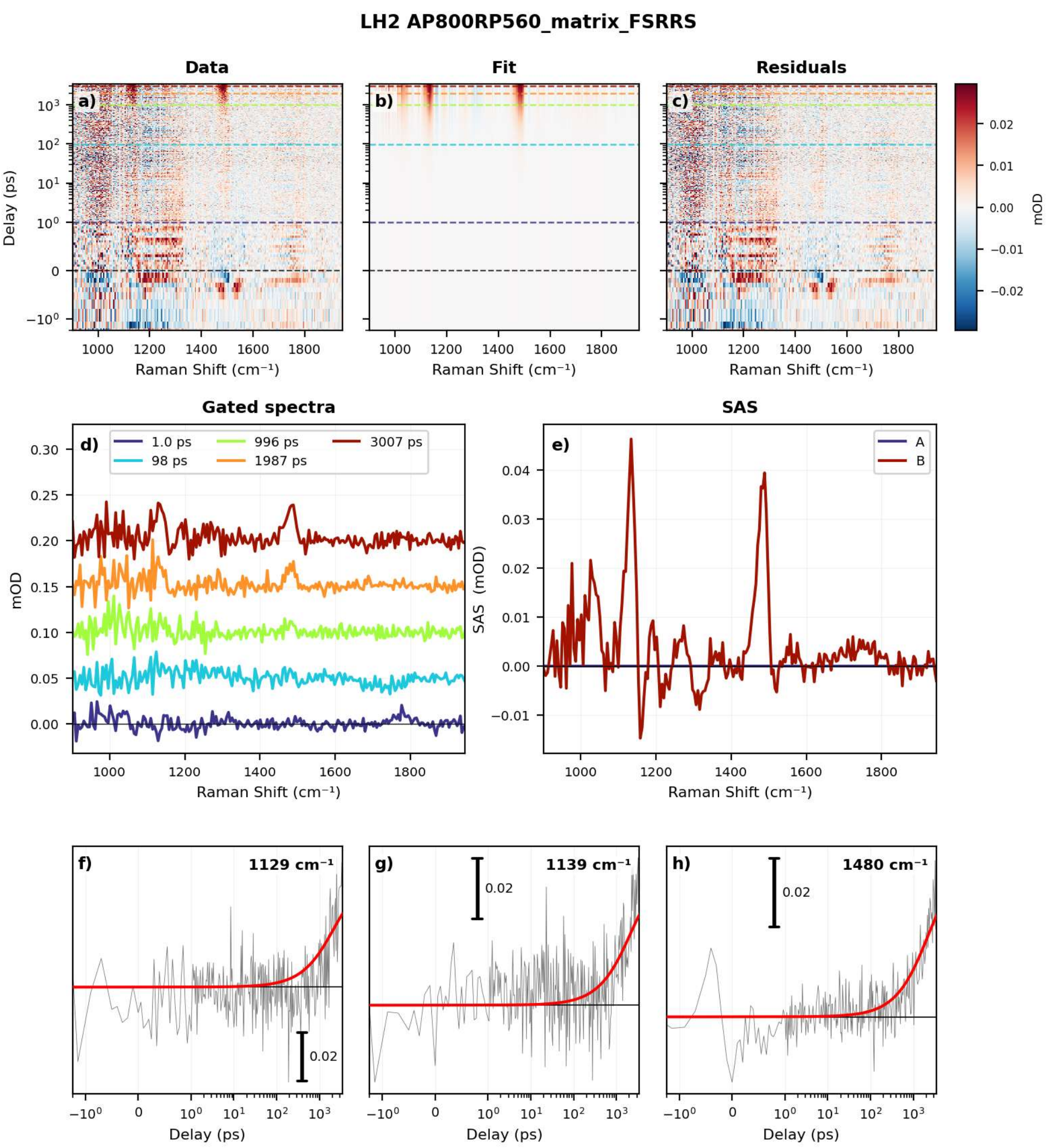

**Figure S17 |** FSRRS of LH2, BChl *a* excitation, actinic pump 800 nm, Raman pump 560 nm. (a) Raw Raman-gain map as a function of Raman shift (cm$^{-1}$) and pump-probe delay (ps). (b) Global sequential fit of the dataset in (a) with a six-component kinetic model. (c) Fit residuals. (d) Time-gated Raman spectra at selected pump-probe delays. (e) Evolution-associated difference spectra (EADS) with the associated component lifetimes. (f-i) Kinetic traces (gray line) of selected vibrational modes, with the sequential-model fit overlaid (red line).

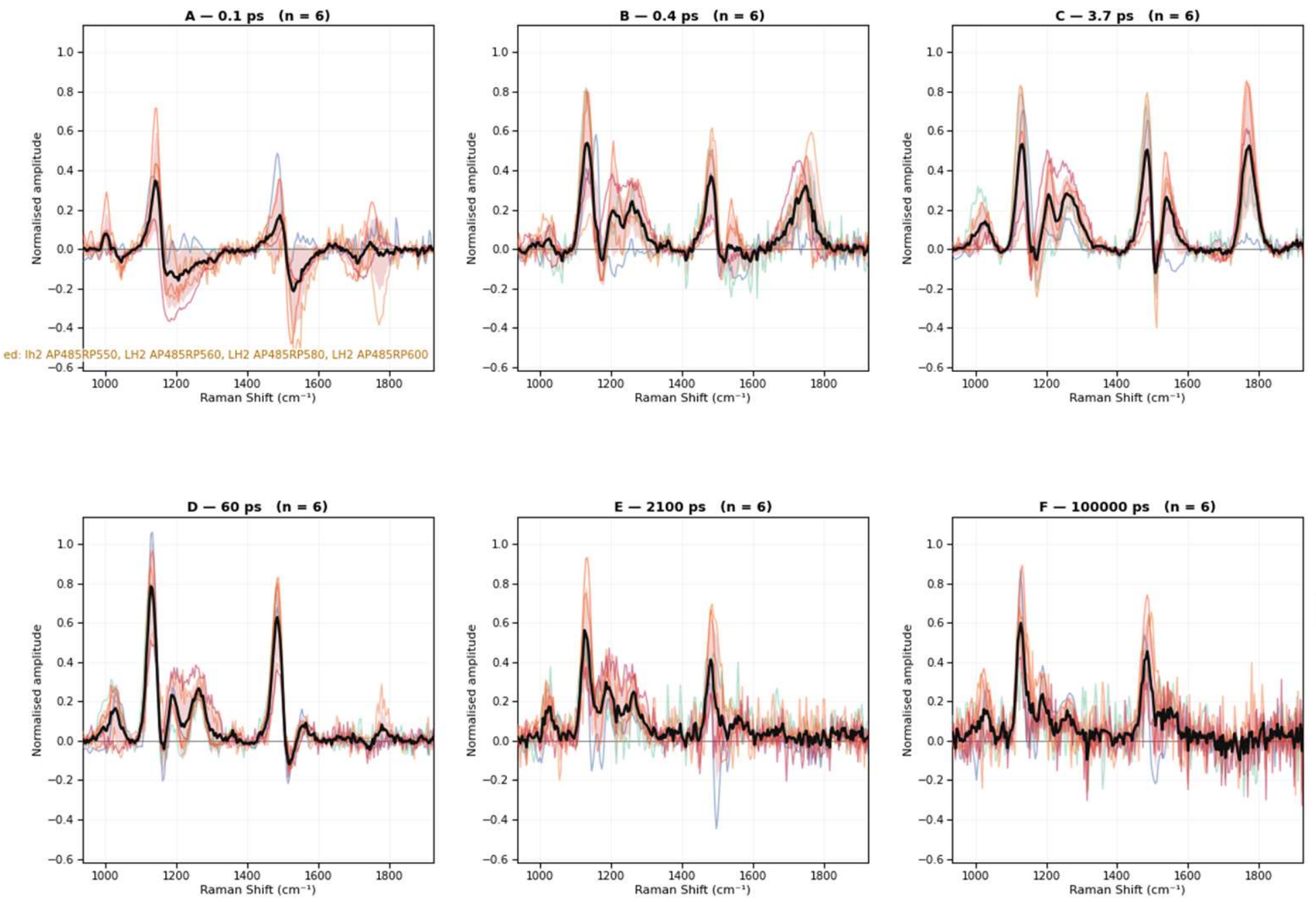

**Figure S18 |** EADS from the multi-matrix global fit of all LH2 datasets. Representative EADS for the six-component sequential model applied simultaneously across carotenoid- and BChl *a*-excitation matrices; the shaded band shows the standard deviation.